\documentstyle[twocolumn,aps,prl,epsf]{revtex}
\begin{document}


\begin{center}
{\large {\bf Universality Classes in Folding Times of Proteins}}
\end{center}

\begin{center}
{\bf
{\sc Marek Cieplak$^1$, and Trinh Xuan Hoang$^2$
}}
\end{center}

\vspace*{1.5cm}
$^1$ Institute of Physics, Polish Academy of Sciences, 02-668
Warsaw, Poland\\
$^2$ Abdus Salam International Center for Theoretical Physics,
Strada Costiera 11, 34100 Trieste, Italy and INFM
Via Beirut 2-4, 34014 Trieste, Italy\\

\vskip 40pt

\noindent
$^*$Corresponding author: \\
Marek Cieplak,\\
Institute of Physics,\\
Polish Academy of Sciences,\\
02-668 Warsaw, Poland\\
E-mail mc@decaf.ifpan.edu.pl\\
Fax 48 22 843 0926


\vskip 30 pt
\noindent { keywords: protein folding; scaling; contact order; Go model;
molecular dynamics; chirality}

\begin{abstract}
Molecular dynamics simulations in simplified models allow one to
study the scaling properties of folding times for many proteins together
under a controlled setting. We consider three variants of the Go models
with different contact potentials and demonstrate scaling described
by power laws and no correlation with the relative contact order parameter.
We demonstrate existence of at least three
kinetic universality classes which are correlated with the types
of structure: the $\alpha$-, $\alpha$--$\beta$-, and $\beta$- proteins
have the scaling exponents of about 1.7, 2.5, and 3.2 respectively.
The three classes merge into one when the contact range is truncated
at a 'reasonable' value. We elucidate the role of the potential
associated with the chirality of a protein.
\end{abstract}

\begin{center}
{\bf INTRODUCTION}
\end{center}

How do size and structure of a protein affect its folding
kinetics is an interesting basic issue that has been debated
in recent years. The size can be characterized by the number, $N$,
of the amino acids that the protein is made of. The distribution of
$N$ across proteins stored in the data banks is peaked around $N$=100
(Cieplak \& Hoang, 2000) and all proteins with a large $N$, like titin
($N\approx$ 30 000), consist of many domains.
There must be then a mechanism
that prevents globular proteins from reaching much larger sizes.
We have argued (Cieplak \& Hoang, 2000) that this is provided by the function
of the protein which requires adoption of a specific conformation.
Folding into it becomes increasingly difficult when $N$ becomes
larger and larger. The sizes of proteins are substantially smaller
than those of the DNA molecules whose
coding function does not depend on the shape.
The native structure of a protein, on the other hand, is believed to be
a decisive factor in its folding mechanism 
(Baker, 2000; Takada, 1999).
\\

A simple parameter that is used to characterize the structure of the
protein is the relative contact order, CO, (Plaxco et al., 1998) defined as average
sequence distance between two aminoacids that interact with each
other, i.e. form a contact, in the native state:
\begin{equation}
CO\;=\;\frac{\sum_{i<j-1}\,\Delta _{ij}\, |i-j|}{N\sum_{i<j-1}\,\Delta _{ij}} \;\;,
\end{equation}
where $\Delta _{ij} $ is 0 if the amino acids $i$ and $j$ do not form
a contact and 1 otherwise.
The relative contact order parameter is  small for $\alpha$-
proteins in which all secondary structures consist of the $\alpha$-helices
because the hydrogen bonds in the helices correspond to $|i-j|=4$.
On the other hand, $\beta$-proteins tend to have larger CO because
the $\beta$ strands that form a sheet often involve amino acids
which are quite distant along a sequence.\\

In their seminal 1998 paper (Plaxco et al., 1998) (paper I),
Plaxco, Baker, and Simons have
argued that folding rates correlate with CO but do not with $N$.
Their argument was based on analysing experimental data on short
proteins that were available in the literature. Their conclusion
was reinforced in the 2000 paper (Plaxco et al., 2000) (paper II)
by Plaxco, Simons, Ruczinski, and Baker in which the
compilation of the kinetic data involved a larger set of proteins,
including those that were considered in paper I.
The later data were also restricted to a much narrower temperature range
of between 20 and 25 $^o$C. Their results for the folding times
(i.e. the inverses of the folding rates) are represented in Figure 1
as a function of $N$ (on the logarithmic scale).
For the purpose of
further discussion, we have divided the data into three classes:
$\alpha$-proteins, $\beta$-proteins, and $\alpha$--$\beta$-proteins.
The $\alpha$-proteins are easily seen to be the fastest folders
but clearly all of the data points are scattered all over the plane
of the figure.\\

This random looking pattern of the data may,
however, be only apparent since the plot might involve mixing distinct
classes of proteins that perhaps should not be compared together.
Figure 2 indeed hints at such a possibility
as the splitting into the $\alpha$-, $\beta$-, and $\alpha$--$\beta$-
structural classes reveals some patterns.
These patterns are shown in different time windows -- the $\alpha$-proteins
are in the window of much shorter  times. There is a growing trend
for the $\beta$- proteins and, if one disregards one outlayer, also
for the $\alpha$--$\beta$-proteins. The data for the $\alpha$-proteins,
however, are puzzling since if they do show an overall trend then
it would be downwards, i.e. the bigger the $N$, the shorter the folding
time which defies a common simple expectation to observe the opposite.\\

The combined data show a strong correlation with the CO parameter.
When the data are split into the three structural classes, as shown
on the right-hand side of Figure 2,
then the correlation remains strong for the
$\alpha$- and $\alpha$--$\beta$-proteins. However, in the crucial
test case of the $\beta$-proteins (the right-bottom panel of Figure 2)
four proteins have nearly the same CO and yet substantially
different folding times.
Thus there are some unsettling issues in our understanding
of the experimental data that would be desirable to solve.\\

Theoretical modeling in simplified models,
despite its well known general shortcomings, is expected to be a tool
of help to identify possible trends because a unified approach
can be applied to many different proteins.
In this paper, we consider 51 proteins: 21 of the $\alpha$--$\beta$
kind with $N$ between 29 and 162, 14 of the $\alpha$-proteins with $N$
between 35 and 154, and 16 $\beta$-proteins with $N$ between 36 and 124.
This set contains the 21 proteins, used in Figures 1 and 2,
that were considered by Plaxco et al. All of the 51 proteins are modelled
in three different ways and studied by the techniques of molecular
dynamics. Even though the three models are all coarse grained
and of the Go type (Abe \& Go, 1981; Takada, 1999) they have very different
kinetic and equilibrium properties when used for a particular
protein. The variations between the models do lead to some differences
in scaling properties of certain parameters, such as the
temperature of the fastest folding, $T_{min}$, or the thermodynamic
stability temperature, $T_f$, but they all agree on a power law
dependence of the folding time, $t_{fold}$, on $N$
\begin{equation}
t_{fold} \sim N^{\lambda}
\end{equation}
and on the lack of any correlation of $t_{fold}$ with CO.
The problems with the experimental results on the $N$ dependence
may be related to the lack of the temperature optimization. The
folding time often depends on the temperature, $T$, and making
choices on the temperature to study kinetics may affect
the outcome of the measurement.
We argue that a demonstrable
trend might arise when all data are collected at $T_{min}$ which
needs to be determined for each protein individually.\\

The theoretically derived lack of correlation of $t_{fold}$ with CO
seems to be a more difficult issue. One may just dismiss it as
characterizing not real life but an approximate model.
On the other hand, the essence of the Go models is that they
are based on the native topology. Thus if such geometry
sensitive models do not 'care' about the contact order then
what models would? We leave it as an open question and this
paper may be just considered to be a report on what are the
properties of three different Go-like models.
Notice, however, that once a Go model is constructed
its contacts are well defined and the kinetics are studied in the context
of such a definition whereas assignment of contacts in experimental
systems is subjective.
It should be pointed out that the contact order in the Go models
is actually quite important but not for the overall folding time --
it is the primary factor that governs the succession of events
during folding 
(Unger, 1996; Hoang \& Cieplak, 2000a; Hoang \& Cieplak, 2000b;
Cieplak et al., 2002a; Cieplak et al., 2002b; Erman, 2001).
In other words, what is important for folding of a protein
is the full "spectrum" of the relevant values of the
sequence distances $|i - j|$ and not just their average value.
A similar point has been argued within a host of models in references
(Galzitskaya and Finkelstein, 1999; Alm and Baker, 1999;
Munoz and Eaton, 1999; Du et al., 1999; and Plotkin and Onuchic, 2000).\\

The power law dependence described by eq. 2 has been proposed
by Thirumalai (Thirumalai, 1995) and then demonstrated
explicitly for several types of lattice models
(Gutin et al., 1996; Zhdanov, 1998; Cieplak et al., 1999).
On the other hand, a number of theories and a recent simulation
of 18 proteins (away from the optimal folding condition)
by Koga and Takada (Koga \& Takada, 2001) suggest a power law dependence for
barrier heights on $N$ and hence an exponential dependence of
$t_{fold}$ on $N$ (Takada \& Wolynes, 1997; Finkelstein \& Badredtinov, 1997;
Wolynes, 1997). Thus
the issue of scaling remains unsettled not only experimentally
but also theoretically. Recently (Cieplak \& Hoang, 2001)
we have demonstrated the power law dependence for one variant (of the
three studied here) Go-like models when applied to 21 proteins which
were mostly of the $\alpha$--$\beta$ kind. The resulting exponent
$\lambda$ turned out to be equal to $2.5 \pm 0.2$.
In this particular variant of the Go model, the
native contact interactions were restricted to a cut-off value of $7.5 \AA$
and the contact potential was described by the
Lennard-Jones form.\\

Here, we extend such studies to the other two kinds of proteins, $\alpha$
and $\beta$, and arrive at a similar value of $\lambda$.
However, when the model is made significantly more realistic by
considering the range of the
native contact interactions as a variable quantity,
then we arrive at a richer picture.
We show that the three classes of tertiary structures also
correspond to three different kinetic universality classes.
The $\alpha$-proteins come with $\lambda$ of around 1.7
(the result obtained previously (Cieplak \& Hoang, 2001) for decoy helical
structures), the $\beta$-proteins are characterized by $\lambda$ close to
3.2, and the $\alpha$--$\beta$-proteins have $\lambda$ near 2.5.
These values do not depend on whether the contact potential
are Lennard-Jones or of the 10--12 form so they are truly
a reflection of the native topology. The power law trends
are pretty evident when the folding times are determined at
$T_{min}$ but harder to see otherwise.
In these studies, the range of the contact interactions has been determined
based on the van der Waals radii of the atoms (Tsai et al., 1999).
Another realistic item
that we implement is the chirality potential -- a term which is
responsible for folding to a conformation of the correct
native chirality. This term affects the kinetics but we show it
not to affect values of the exponent $\lambda$.\\

The growth of $t_{fold}$ with $N$ indicates increasingly deteriorating
folding conditions.
Our studies of scaling of $T_{min}$ and $T_f$
indicate that asymptotically $T_f$ becomes substantially lower
than $T_{min}$ which signifies an onset of slow glassy  kinetics
before the system is near the native conformation.
This adds to the deterioration of foldability and suggests
the limitation in the observed values of $N$.
The three models considered here have $T_{min}$ and $T_f$ varying
as a function of $N$ in different ways, though they agree asymptotically.
Among the three models, the Lennard-Jones contact potential
with the variable $R_c$ appears to have the most appealing kinetic
properties in that it leads to a very good foldability for a small $N$.
This should be our simple model of choice in future studies.
However, the issue of the scaling trends needs now to be
studied in models that reach beyond the Go approximation
and in experiments with a protocol that involves optimization.\\

\begin{center}
{\bf MATERIALS AND METHODS}
\end{center}

\underline{A. The Hamiltonian}\\

An input for the construction of the Go model is a PDB file (Bernstein et al., 1977)
with the coordinates
of all atoms in the native conformation. The coordinates are used
to determine the length related parameters of the model. Whereas
all energy and temperature related parameters are expressed
in terms of a common unit -- $\epsilon$. We model 51 proteins.
In addition to the proteins listed in the caption of Figure 1,
we also consider 1cti(29), 1cmr(31), 1erc(40), 1crn(46), 7rxn(52),
5pti(58), 1tap(60), 1aho(64), 1ptx(64), 1erg(70), 102l(162) which are
of the $\alpha$--$\beta$ type, or unstructured, then
1ce4(35), 1bba(36), 1bw6(56), 1rpo(61), 1hp8(68), 1ail(73), 1ycc(103)
which are of the $\alpha$ type, and
1cbh(36), 1ixa(39), 1ed7(45), 1bq9(53), 2cdx(60), 2ait(74), 1bdo(80),
1wit(93), 1who(94), 6pcy(99), 1ksr(100), 4fgf(124) which are
of the $\beta$ type. The symbols are the PDB codes and the numbers
in brackets indicate the corresponding value of $N$.
The choice of these proteins was motivated by their size
but otherwise random.\\

We consider several variants of the Go models. In each case, the Hamiltonian
consists of the kinetic energy and of the potential energy,
$E_p(\{\bf r_i\})$,
which is given by
\begin{equation}
E_p(\{{\bf r_i}\})\;=\; V^{BB} \;+\;V^{NAT} \;+\;V^{NON} \;+\;V^{CHIR} \;\;.
\end{equation}
The first term, $V^{BB}$ is the harmonic potential
\begin{equation}
V^{BB} = \sum_{i=1}^{N-1} \frac{1}{2}k (r_{i,i+1} - d_0)^2 \;\;,
\end{equation}
which tethers consecutive beads at the equilibrium bond length,
$d_0$, of $3.8\AA$. Here,
$r_{i,i+1}=|{\bf r}_i - {\bf r}_{i+1}|$ is the distance between
the consecutive beads and $k=100 \epsilon /$\AA$^2$, where
$\epsilon$ is the characteristic
energy parameter corresponding to a native contact.\\

The native contacts are defined either through the distances between
the C$^\alpha$ atoms or through an all-tom consideration.
The first choice, used by us previously 
(Hoang \& Cieplak, 2000a; Hoang \& Cieplak, 2000b; Cieplak \& Hoang, 2001),
is to take a uniform cut-off distance,
$R_c$, of $7.5\AA$, below which a contact is said to be present.
In the second choice, used here in most cases, all the heavy atoms
present in the PDB file are taken into account.
Specifically, a pair of aminoacids is considered to form a contact
if any pair of their non-hydrogen atoms have a native separation which
is smaller than 1.244 $(R_i+R_j$), where $R_i$ are the van der Waals
radii of atom $i$, as listed in ref. (Tsai et al., 1999). This critical
separation corresponds to the point of inflection of the Lennard-Jones
potential. Figure 3 shows the distribution of the effective
contact ranges as obtained for an $N$=162 protein T4 lysozyme
with the PDB code 102l which consists of 10 $\alpha$-helices and
3 $\beta$-strands. There are 339 native contacts in this case and they
range in value between 4.36 and 12.80 $\AA$. It is clear that truncating
this distribution at whatever "reasonable" value, which is often
taken to be in the range between 6.5 and 8.5 $\AA$ would result
in a substantial removal of the relevant interactions. Thus insisting
on a uniform cutoff value is expected to have noticeable dynamical
effect.\\

We consider two variants of the interactions in the native contacts.
The first variant is the 6--12 Lennard-Jones potential
\begin{equation}
V^{NAT}_{6-12} =
\sum_{i<j}^{NAT}4\epsilon \left[ \left( \frac{\sigma_{ij}}{r_{ij}}
\right)^{12}-\left(\frac{\sigma_{ij}}{r_{ij}}\right)^6\right],
\end{equation}
where the sum is taken over all native contacts.
The parameters $\sigma_{ij}$ are chosen so that each contact in the native
structure is stabilized at the minimum of the potential,
and $\sigma \equiv 5\AA$ is a typical value.
The second variant is the 10--12 potential
\begin{equation}
V^{NAT}_{10-12} =
\sum_{i<j}^{NAT}\epsilon \left[ \left( 5\frac{r^{(n)}_{ij}}{r_{ij}}
\right)^{12} - 6 \left(\frac{r^{(n)}_{ij}}{r_{ij}}\right)^{10}\right],
\end{equation}
where $r^{(n)}_{ij}$ coincides with the native distance. This potential
is frequently used to describe hydrogen bonds (Clementi et al., 2000).
For each pair of interacting amino acids, the two potentials
have a minimum energy of $-\epsilon$ and are cut off at $20 \AA$.
The non-native interactions, $V^{NON}$, are purely repulsive and are
necessary to reduce the effects of entanglements.
They are taken as
the repulsive part of the Lennard-Jones potential that corresponds
to the  minimum occurring at $5\AA$. This potential is truncated
at the minimum and shifted upward so that it reaches zero energy
at the point of truncation.\\

The final term in the Hamiltonian takes into account the chirality.
Natural proteins have right handed helices but a Go model as
described above involves chiral frustration: one end of a helix
may want to fold into a right handed helix and another into a left
handed one and "convincing" one end to agree with the twist of the other
takes time and delays folding. Such a
frustration would not arise naturally. In order to prevent it,
we add a term which favors the native sense of the overall chirality
at each location along the backbone.
A chirality of residue $i$ is defined as
\begin{equation}
C_i = \frac{\left( {\bf v}_{i-1} \times {\bf v}_{i} \right)
\cdot {\bf v}_{i+1}}{d_0^3},
\end{equation}
where ${\bf v}_{i}={\bf r}_{i+1} - {\bf r}_{i}$.
A positive $C_i$ corresponds to right-handed chirality. Otherwise the
chirality is left-handed. The values of $C_i$ are essentially
between $-1$ and $+1$. The distribution of $C_i$ in
21 $\alpha$--$\beta$-proteins considered in this study is shown
in Figure 4. It is seen to be bimodal. The values
in the higher peak correspond to locations within the helical
secondary structures.
The chiral part of the Hamiltonian is then given phenomenologically by
\begin{equation}
V^{CHIR} =
\sum_{i=2}^{N-2}\frac{1}{2} \; \kappa \; C_i^2 \; \Theta ( - C_i^{NAT} ),
\end{equation}
where $\Theta$ is the step function (1 for positive arguments
and zero otherwise),
$C_i^{NAT}$ is the chirality
of residue $i$ in the native conformation,
and $\kappa$ is taken, in most cases, to be equal to $\epsilon$.
However, a criterion for selection of its proper value
remains to be elucidated.
The idea behind this
particular form of $V^{CHIR}$ is that when the local chirality
agrees with the native chirality then there is no effect on the
energy. On the other hand, a disagreement in the chirality is
is punished by a cost which is quadratic in chirality.\\

$V^{CHIR}$ has the strongest effect
on the helical structures. However, it affects the sense of a twist of the
whole tertiary structure.
The chirality term enhances the dynamical bias towards the native
structure during the folding process and helps avoiding non-physical
conformations such as left-handed helices.
$V^{CHIR}$ is a four-body potential.
In this respect this term is similar
to potentials that involve dihedral angles 
(Veitshans et al., 1997; Clementi et al., 2000; Settanni et al., 2002).
The dihedral terms enhance stability of a model of the protein
but usually have no bearing on the chirality (Veitshans et al., 1997) unless
they involve directly the values of native dihedral angles
(Clementi et al., 2000; Settanni et al., 2002).
\\

\underline{B. The time evolution}\\

The time evolution of unfolded conformations to the native state is simulated
through the methods of molecular dynamics as described in details
in (Hoang \& Cieplak, 2000a; Hoang \& Cieplak, 2000b)
(see also (Cieplak et al., 2002a; Cieplak et al., 2002b) in the
context of the Lennard-Jones contact potentials.
The beads representing the amino acids
are coupled to Langevin noise and damping terms to
mimic the effect of the surrounding solvent and provide thermostating
at a temperature $T$.
The equations of motion for each bead are
\begin{equation}
m\ddot{{\bf r}} = -\gamma \dot{{\bf r}} + F_c + \Gamma \ \ ,
\end{equation}
where $m$ is the mass of the amino acids represented by each bead.
A similar approach in the context of proteins has also been adopted
in references (Guo and Thirumalai, 1996; Berriz et al., 1997; and
Eastman and Doniach 1998).
The specificity of masses has turned out to be irrelevant
for kinetics (Cieplak et al., 2002a) and it is sufficient to consider
masses that are uniform and equal to the average amino acidic mass.
$F_c$ is the net force due to the molecular potentials and external forces,
$\gamma$ is the damping constant, and
$\Gamma$ is a Gaussian noise term with dispersion
$\sqrt{2\gamma k_B T}$.
For both kinds of the contact potentials, time is measured in units of
$\tau \equiv \sqrt{m \sigma^2 / \epsilon}$, where $\sigma$ is $5\AA$.
This corresponds to the
characteristic period of undamped oscillations at the bottom
of a typical 6--12 potential.
For the average amino acidic mass and $\epsilon$ of order 4kcal/mol,
$\tau$ is of order 3$ps$.
According to Veitshans et al. (Veitshans et al., 1997),
realistic estimates of damping by the solution correspond to a
value of $\gamma$ near 50 $m/\tau$.
However, the folding times have been found to depend on
$\gamma$ in a simple linear fashion for $\gamma > m/\tau$
(Hoang \& Cieplak, 2000a; Hoang \& Cieplak, 2000b; Klimov \& Thirumalai, 1997).
Thus in order to
accelerate the simulations, we work with $\gamma \;= \; 2 m/\tau$
but more realistic time scales are obtained when the folding times
are multiplied by 25.
The equations of motion are
solved by means of the fifth order Gear predictor-corrector
algorithm (Gear, 1971) with a time step of $0.005\tau$.
\\

The magnitude of the viscous effects, as controlled by the parameter
$\gamma$, has to be sufficiently large so that the scenarios of the folding 
events are not dominated by the inertial effects. Otherwise the
scenarios would depend on the spacial and not on the sequencial separation
between the amino acids. Figure 5,
for crambin as an illustration,  shows that even though our value of
$\gamma$ of 2 is reduced compared to the values that are expected to be
realistic it already corresponds to sufficiently strong damping 
with the minimal inertial effects. Figure 5 gives average first times 
needed to establish contacts separated by the sequence length $|i-j|$
for three values of $\gamma$: 2, 12, and 24 $m/\tau$. To the leading order,
the times to establish the contacts (and also the folding times)
are linear functions of $\gamma$ so one can show them together
by proper rescaling. Furthermore, the whole pattern of the events
is insensitive to the value of $\gamma$.
Starting with this figure, we adopt the convention that the symbol
sizes give measures of the error bars in the quantity that is plotted.\\

The folding time is calculated as the median first passage time, i.e.
the time needed to arrive in the native conformation from an
unfolded conformation.
It is estimated based on between 101 and 201 trajectories.
$T_{min}$ is defined as a temperature
at which $t_{fold}$ has a minimum value when plotted vs. $T$.
For small values $N$, the U-shaped dependence of $t_{fold}$ on $N$
may be very broad and then $T_{min}$ is defined as the
position of the center of the U-shaped curve.
The simplified criterion for an arrival in the native conformation
to be declared
is based on a simplified approach
in which a protein is considered folded if all beads that form a native
contact are within the cutoff distance of $1.5 \sigma _{ij}$
or $1.2 r^{(n)}_{ij}$ for the 6--12 and 10--12 potentials respectively.\\

The stability temperature $T_f$ is determined through the
nearly equilibrium calculation of the probability that the protein
has all of its native contacts established. $T_f$ is the temperature at
which this probability crosses $\frac{1}{2}$.  The calculation is based
on least 5 long trajectories that start in the native state in order to
make sure that the system is in the right region of the conformation
space. 
It should be noted that, in the literature,
the frequently used estimate of 
the folding temperature is determined 
through the position of the maximum in the specific heat.
This yields a $T_f^{'}$ which is typically larger than $T_f$. 
Our probabilistic
interpretation has the disadvantage of being dependent on the precise
definition of what constitutes the native basin (and thus only the 
approximate location of $T_f$ is of relevance) but it has the advantage
of relating only to the native basin and not to any other 
valleys in the phase space. 
In most of our systems, $T_f$ is found to be comparable
to $T_{min}$, while both of them are always lower than $T_f^{'}$.
Furthermore, in most cases, even though when $T_{min}$ is found to
be higher than $T_f$, the folding times at $T_f$ are comparable to those 
at $T_{min}$ which indicates that the model is unfrustrated in the
conventional sense. Only in some very few cases, the folding times
at $T_f$ are excessively long to be determined in our simulation.
This behaviour
probably corresponds to a structural frustration (Clementi et al., 2000) 
embedded in the native conformation.
\\

An alternative to the contact-based criterion for folding is to provide
a more precise delineation of the native basin as in 
ref. (Hoang \& Cieplak, 2002b)
or relate the criterion to a cutoff in the value of the RMSD distance
away from the native conformation. These approaches are illustrated in
Figure 6 which shows the dependence of the folding time, $t_{fold}$,
vs. $T$ for a synthetic  $\alpha$-helix 
(H16 of reference (Hoang \& Cieplak, 2002a))
and $\beta$-hairpin (B16 of the same reference) that both consist
of 16 monomers. Whichever criterion for folding is used, the
folding curves are U-shaped and the  non-zero chirality term
extends the region
of the fastest folding both towards the low and high temperature ends.
For the hairpin, the effect is smaller but still clearly present.\\

When it comes to model proteins, we used only the contact-based folding
criterion. An illustration of the role of the chirality potential
is provided in Figure 7 for crambin ($N$=46, the PDB code 1crn) which
is a protein of the $\alpha$--$\beta$ type. The top panel, for $R_c=7.5 \AA$,
shows that the shortest time of folding is somewhat reduced by $V^{CHIR}$
but the biggest impact is on the range of temperatures at which folding
is optimal, almost by the factor of 2, especially in the low $T$ regime.
For the $\beta$ proteins, the effect of the chirality potential is generally
smaller. For the SH3 domain coded 1efn the change due to
$V^{CHIR}$ is hard to detect (not shown) but for the I27 globular domain of
titin, coded 1tit, it is quite substantial on the low $T$ side of the
curve (Figure 8). We conclude
that incorporation of the chirality term in the Hamiltonian
appears to reduce structural frustration in these models and thus
makes the models more realistic. For all of the results presented here
from now on (except for Figure 12), the chirality term is included.\\

Another simple way to enhance the realism of the Go models is suggested
by Figure 3: calculate the range of the contact potential instead
of taking one uniform cutoff value. When we compare the case of the
Lennard-Jones contact potential with the uniform or variable $R_c$
then the nature of the  effect on the kinetics strongly depends
on the protein. For instance, for
the protein 1crn (Figure 7, bottom panel) there
is essentially no difference. 
On the other hand, a dramatic
narrowing of the U-curve is observed for 1tit (Figure 8).\\

On switching the 6--12 potential to the 10--12 potential all of the
kinetic U-curves become substantially narrower (Figures 7 and 8).
This is related to the fact that the potential well corresponding
to the 10--12 potential is narrower which makes folding a task
that requires more precision. Note, that the two potentials have the
same energy ($-\epsilon$) at the minimum so the temperature
scale are comparable.\\

We have demonstrated that there are many ways to construct variants
of the Go models and they all come with distinctive folding characteristics.\\

\begin{center}
{\bf RESULTS}
\end{center}

\underline{A. The 6--12 potential with the variable contact range}\\

Figure 9 shows the median values of $t_{fold}$ at $T_{min}$ for
the Lennard-Jones contact potential when the presence of the
native contact is determined through the van der Waals sizes
of the atoms (and with the chirality term included).
Figure 9 data divides the data into the
three structural classes. There are a few outlayers (one is
the 1aps protein which appears to be a poor folder also
experimentally) but basically there are clear linear trends
on the log-log scale which indicates validity of the power law,
eq. (2). The values of the exponents 1.7 for the $\alpha$-proteins
and 3.2 for the $\beta$-proteins agree with those found for
decoy structures (Cieplak \& Hoang, 2001). The decoy structures were
constructed from homopolymers and the contact range was not variable
due to the lack of atomic features in the decoys.
Figure 10 replots the same data together
to indicate that the trends identified in the classes are
identifiably distinct. Thus the structural classes also
correspond to the kinetic universality classes.\\

Figure 11 shows data equivalent to those on Figure 9 but now
the folding times are determined at $T_f$, as an example of a
situation that may be encountered away from the optimal conditions.
The data points  show a much larger scatter away form the
trend identified at $T_{min}$. The optimal trend seems still
dominant but it is so much harder to see. This should be analogous
to results obtained experimentally.\\

It is interesting to figure out what is the effect of the chirality
potential on the scaling results. Figure 12 refers to the
$\alpha$-proteins and it compares the case of $\kappa=0$ to $\kappa=\epsilon$.
Proteins with small values of $N$ are not sensitive to the value of $\kappa$
but for $N > \approx 50$ taking the chirality into account
accelerates the kinetics quite noticeably. The 'asymptotic' scaling
behavior remains unchanged -- the exponent $\lambda$ of 1.7 is valid
for both cases, though a somewhat larger value for $\kappa$=0
cannot be ruled out (but certainly not as large as 2.5).
We have checked that the data points for $\kappa=2\epsilon$, though
corresponding to a bit faster times than for $\kappa=\epsilon$,
are in practice indistinguishable from the latter in the scale of the figure.
This observation suggests a behavior which saturates with a growing $\kappa$.\\

As pointed out in Ref. (Cieplak et al., 1999), the dependence of $T_f$
and $T_{min}$ on $N$ may offer additional clues about
the foldability at large $N$. Figure 13 suggests that the
$\alpha$- and $\alpha$--$\beta$-proteins are excellent folders
for small values of $N$ since then $T_{min}$ is less than $T_f$.
$T_f$ appears to have no systematic trend with $N$ but
the data for $T_{min}$ suggest a
weak growth, approximately proportional to $log (N)$.
Around $N$ of 50 the trend associated with $T_{min}$ crosses
the average value of $T_f$ and from now on $T_f$ is lower than $T_{min}$.
This suggests that  asymptotically the energy landscape of the system
would be too  glassy-like to sustain viable folding. Thus
accomplishing folding  would require breaking into independently
folding domains domain or receiving an external assistance,
e.g. from chaperons whereas our studies are concerned with
individual proteins. Figure 13 also suggests that the $\beta$
proteins behave somewhat differently since they exhibit no trend in
$T_{min}$ in the range studied and already for small values of $N$
$T_{min}$ exceeds $T_f$. Nevertheless the differences between
the three structural classes are minor because they all show
a border line behavior: the proteins in the range up to $N$=162
are not excellent but just adequate folders, at least in this model.\\

It is interesting to point out that neither $t_{fold}$ nor the
characteristic temperatures indicate any demonstrable correlation
with the relative contact order defined in eq. 1.
This is shown in Figure 14: for a given value of CO we find systems
both with long and short folding times or both high and low
values of $T_{min}$.\\

\underline{B. The 10--12 potential with the variable contact range}\\

We now check the stability of our results against the change
in the form of the contact potential with the same characteristic
energy scale. Figure 15 shows that when the Lennard-Jones potential
is replaced by the 10--12 potential, with keeping all other
Hamiltonian parameters intact, the scaling trends for $T_{fold}$
are consistent with those displayed in Figure 9 and confirm the
existence of the three universality classes.\\

Figure 16 suggests that the 10--12 systems are also border line
in terms of the positioning of $T_{min}$ vs $T_f$ but the weak
growing trends for the $\alpha$- and $\alpha$--$\beta$-proteins
are gone. The lack of correlations with the relative contact order also
holds for the 10--12 potential (not shown).\\

\underline{C. The 6--12 potential with $R_c=7.5\AA$}\\

We now return to the Lennard-Jones potential and make the drastic,
as evidenced by Figure 3, change that only those native contacts
are considered whose range does not exceed $7.5\AA$.
The resulting data are shown in Figure 17. The top panel indicates
that $\lambda$ of about 2.5 is still consistent with the trend
obtained. However, $\lambda$ of 1.7 is quite off the mark for the
$\alpha$-proteins. The exponent of 3.2 for the $\beta$-proteins
is not ruled out but the scatter in the data points is bigger
than in the bottom panel of Figure 9. Taken together with the
results for the $\alpha$-proteins, the most likely conclusion
is that the fixed, and invasive, cut off in the contact range
looses the ability to distinguish between the structural classes
and all such  models of the proteins would be characterized
by a single exponent $\lambda$ of 2.5 as found in ref. (Cieplak \& Hoang, 2001).
This is illustrated in Figure 18 where the data corresponding
to various structural classes are displayed together. They seem
to be consistent with just one trend.\\

Figure 19 shows $T_{min}$ and $T_f$ for the case with $R_c=7.5\AA$.
It suggests that among the three models studied here, the one
with the cut off in the contact range is the worst kinetically
because the gap between the band of values of $T_{min}$
and the band of values of $T_f$ is the largest. This indicates
that precise values of the contact range are important in the task
of putting pieces of a protein together in the folding process.
Also in this model, there is no correlation with the relative contact
order parameter.\\

\begin{center}
{\bf DISCUSSION}
\end{center}

We have studied 3 variants of the Go model through the molecular
dynamics simulations and demonstrated the power law dependence
of the folding time on $N$ and lack of dependence on CO.
Furthermore, the models with the variable contact range allow
one to identify (at least) three kinetic universality classes
corresponding to three different values of the exponent $\lambda$.
The lowest exponent found for the $\alpha$- structures is consistent
with the widely held belief that the $\alpha$-helices are structures
that are optimal kinetically 
(Micheletti et al., 1999; Maritan et al., 2000).
The scaling behavior of $T_{min}$ and $T_f$, taken together
with the increasing $t_{fold}$ suggests an asymptotic
emergence of a glassy behavior.
As a technical improvement, we have highlighted benefits
of introducing the chirality potential.\\

Recently,
Koga and Takada (Koga \& Takada, 2001) have also studied scaling of $t_{fold}$
in proteins approximated by the Go model.
They have considered the 10--12 potential
that was augmented by potentials which involved the dihedral angles
(but no chirality).
They have determined the folding temperature
through the maximum in the specific heat.
Their studies at $T_f^{'}$, done for 18 proteins with $N$ in the range
between 53 and 153, suggest a $t_{fold}$ that
exponentially depends on the relative contact order multiplied by $N^{0.6}$.\\

It is thus interesting to check on this conclusion in the framework
of our approach. Figure 20
shows $log(t_{fold})$ vs. CO$\times N^{0.6}$ for our best model,
i.e. for the Lennard-Jones contact potential with variable
contact range.
It is clear that the data at $T_{min}$ (the left panels)
show significantly less scatter
than at $T_f$ (the right panels)
so the distinction between the power law and the
exponential function is certainly not due to considering
different temperatures. Figure 20 does suggest a correlation with
CO$\times N^{0.6}$ (the data plotted vs. $N^{0.6}$ without the
CO factor have a similar appearance indicating the irrelevance of CO
in such theoretical studies)
and Koga and Takada quote a correlation
level of 84\% for their data.
It is not very easy to distinguish
between the power law and the exponential dependencies without
a significant broadening of the range in the values of $N$.
Figure 21 shows the data of Figure 9 redisplayed on the log - linear scale.
The exponential trends, $t_{fold} \sim exp(N/\xi)$, cannot be ruled
out and the correlation levels are 75\%, 94\%, and 95\% for the
$\alpha$--$\beta$, $\alpha$, and $\beta$ structural classes
respectively whereas the corresponding values for the log-log
plots are 81\%, 97\%, and 94\%. Even though the power law fits
appear better (or, in the case of the $\beta$ proteins about the same) the
important point is that the exponential fits also suggest
existence of the three different kinetic universality classes
since the characteristic values of the $\xi$ parameter, as displayed
in the Figure, are clearly distinct.
Our trends displayed in Figure 9 seem much less scattered than those
shown in Figure 20, especially in the right hand panels of Figure 20.
However, while we argue in favour of the three universality classes
and then the power laws, we see
a need for further studies and better understanding of these issues.\\

It has been found recently (Cieplak \& Hoang, 2002c) that the kinetics of
Go models are very sensitive to the selection of what constitutes the 
proper set of the native contacts. For instance, if one declares
a uniform cutoff range, $R_c$, between the $C^{\alpha}$ atoms
for making a contact, then the dependence of $t_{fold}$ on $R_c$ is
strong and non-monotonic. Koga and Takada declare the contact as 
occurring if two non-hydrogen atoms in a pair of amino acids
are in a distance of less than either $5.5\AA$ or $6.5\AA$
(and it is stated that the results are stable with respect to this 
choice). Our definition of the contacts, on the other hand,
involves the atomic sizes which yields a different contact map
and leads to different folding times.\\

The basic unsolved question is why do the folding times
in various Go models do not depend on the contact order even though
the primary ingredient of any Go model is the geometry of the native
state of a protein. One technical problem
with the contact order is that the very notion of a contact is fairly
subjective. Consider, for instance, the G protein -- the PDB code is
1gb1 for the structure determined by NMR and 1pga for the
crystallographic structure.
When we make use of the van der Waals radii  then we get
$CO=0.239$ for 1gb1 and 0.250 for 1pga.
The alternative procedure
is to consider two residues contacting if they
contain non-hydrogen atoms within a distance of $d$.
For $d$ equal to 3, 4, 5, 6, 7 and 8 $\AA$, our
procedure yields CO of 0.194, 0.220, 0.235, 0.252, 0.277, and 0.295
respectively (for the 1pga structure it is 0.257
if the cutoff of $6 \AA$ is used -- i.e. not very different).
Plaxco et al. (Plaxco et al., 1998; Plaxco et al., 2000)
used the value of $d=6\AA$,
and they quoted CO of 0.173 for this case. The notable difference
from our value arises from the fact that in their calculation
(Plaxco -- private communication),
all of the contacts made by the atoms (i.e. up to dozens for a pair of
amino acids) contribute to the value of CO if the corresponding distance
does not exceed $d$. Furthermore, the 'contacts' between
consecutive residues (i.e. between $i$ and $i+1$) are
taken into account. In our calculation, the shortest local
contacts are of the $i$, $i+2$ type.
Note that the values of CO vary with $d$ quite substantially
(on the scale of the figures
involved) and the value obtained at $d=6\AA$ is about 45\% larger than
that quoted by Plaxco et al.
The important point, however, is not that much
what is the absolute value of CO but whether its
correlation with the folding rate is sensitive to the choice
of a specific definition of CO that is adopted.
We have found that, quite remarkably, this correlation
in the set of the experimentally studied
proteins remains strong even when our procedure for the
calculation of CO is used. We find that even though
the scatter away from the trend is noticeably larger
than when using  the CO$_P$ -- the values of CO quoted by Plaxco et al. --
the correlations with CO remain robust and
some dependence on CO develops in the case of the $\beta$-proteins.
It is hoped that further interactions and iterations between theory
and experiment will  make the issues of size and contact order
dependence more definitive. The notion of universality classes in
proteins should play an important role in this process.\\

\begin{center}
{\bf ACKNOWLEDGMENTS}
\end{center}
We appreciate Kevin Plaxco's help in elucidating his papers to us
and for his comments.
Fruitful discussions with A. Maritan, J. R. Banavar,
M. O. Robbins, and G. D. Rose are greatly appreciated.
MC thanks the Department of Physics and Astronomy at Rutgers University for
providing him with the computing resources.
This research was supported by the grant 2P03
from KBN (Poland).
\\

\begin{center}
{\bf REFERENCES}
\end{center}

\begin{description}

\item
Abe, H., and N. Go.  Noninteracting local-structure model of folding and
unfolding transition in globular proteins. II. Application to two-dimensional
lattice proteins.  1981. {\it Biopolymers} 20:1013-1031

\item
Alm, E., and D. Baker.
Prediction of protein-folding mechanisms from free-energy
landscapes derived from native structures.
1999. {\it Proc. Natl. Acad. Sci. USA} 96:11305-11310

\item
Baker, D. A surprising simplicity to protein folding.  2000. 
{\it Nature} 405:39-42

\item
Bernstein, F. C., T. F. Koetzle, G. J. B. Williams, E. F. Meyer Jr., 
M. D. Brice, J. R. Rodgers, O. Kennard, T. Shimanouchi, and M. Tasumi.  
The Protein Data Bank: a computer-based archival file for 
macromolecular structures.  1977. {\it J. Mol. Biol.} 112:535-542

\item 
Berriz, G. F., A. M. Gutin, and E. I. Shakhnovich.
Cooperativity and stability in a Langevin model of proteinlike folding.
1997. {\it J. Chem. Phys.} 106:9276-9285

\item
Cieplak, M., and  T. X. Hoang.  Kinetics non-optimality and vibrational 
stability of proteins.  2001. {\it Proteins} 44:20-25 

\item
Cieplak, M., and T. X. Hoang.  Scaling of folding properties in Go models 
of proteins.  2000. {\it J. Biol. Phys.} 26:273-294

\item
Cieplak, M., T. X. Hoang, and M. O. Robbins.  Folding and stretching in a 
Go-like model of Titin.  2002a. {\it Proteins} (in press)

\item
Cieplak, M., T. X. Hoang, and M. O. Robbins.  
Thermal folding and mechanical unfolding pathways of protein secondary 
structures.  2002b. {\it Proteins} (in press)

\item
Cieplak, M., and T. X. Hoang.
The range of contact interactions and the kinetics of the Go models
of proteins. 2002c. {\it Int. J. Mod. Phys. C} (in press),
and  http://arXiv.org/abs/cond-mat/0206299

\item
Cieplak, M., T. X. Hoang, and M. S. Li.  Scaling of folding properties in
simple models of proteins.  1999. {\it Phys. Rev. Lett.} 83:1684-1687

\item
Clementi, C., H. Nymeyer, and J. N. Onuchic.  
Topological and energetic factors: what determines the structural details
of the transition state ensemble and ``on-route'' 
intermediates for protein folding? An investigation for small globular 
proteins.  2000. {\it J. Mol. Biol.} 298:937-953 

\item
Du, R., V. S. Pande, A. Y. Grosberg, T. Tanaka, and E. I. Shakhnovich.
On the role of conformational geometry in protein folding.
1999. {\it J. Chem. Phys.} 111:10375-10380

\item
Eastman, P., and S. Doniach.
Multiple time step diffusive Langevin dynamics for proteins.
1998. {\it Proteins} 30:215-227

\item
Erman, B.  
Analysis of multiple folding routes of proteins by a coarse-grained 
dynamics model.  2001. {\it Biophysical J.} 81:3534-3544

\item
Finkelstein, A. V., and A. Y. Badredtinov.  Rate of protein folding near the
point of thermodynamic equilibrium between the coil and the most stable
chain fold.  1997. {\it Fold. Des.} 2:115-121

\item
Galzitskaya, O. V., and A. V. Finkelstein.
A theoretical search for folding/unfolding nuclei in three-dimensional
protein structures.
1999. {\it Proc. Natl. Acad. Sci. USA} 96:11299-11304

\item
Gear, W. C.  {\it Numerical Initial Value Problems in Ordinary  
Differential Equations.} 1971. Prentice-Hall, Inc. 

\item
Guo, Z., and D. Thirumalai.
Kinetics and thermodynamics of folding of a de novo designed four-helix
bundle protein.
1996. {\it J. Mol. Biol.} 263:323-343

\item
Gutin, A. M., V. I. Abkevich, and E. I. Shakhnovich.  
Chain length scaling of protein folding time, 1996. 
{\it Phys. Rev. Lett.} 77:5433-5436 

\item
Hoang, T. X., and M. Cieplak.  
Molecular dynamics of folding of secondary structures in Go-like 
models of proteins.  2000a. {\it J. Chem. Phys.} 112:6851-6862

\item
Hoang, T. X., and M. Cieplak.  
Sequencing of folding events in Go-like proteins.  2000b. 
{\it J. Chem. Phys.} 113:8319-8328 

\item
Klimov, D. K., and D. Thirumalai.  
Viscosity Dependence of the Folding Rates of Proteins.  1997. 
{\it Phys. Rev. Lett.} 79:317-320

\item
Koga, N., and S. Takada.  
Roles of native topology and chain-length scaling in protein folding: 
a simulation study with a Go-like model.  2001. {\it J. Mol. Biol.} 313:171-180

\item
Maritan, A., C. Micheletti, and J. R. Banavar.  
{\it Role of secondary motifs in fast folding polymers: A dynamical 
variational principle.} 2000. Phys. Rev. Lett. 84:3009-3012

\item
Micheletti, C., J. R. Banavar, A. Maritan, and F. Seno.  
Protein structures and optimal folding from a geometrical variational principle.
 1999. {\it Phys. Rev. Lett.} 82:3372-3375

\item
Munoz, V., and W. A. Eaton.
A simple model for calculating the kinetics of protein folding from
three-dimensional structures.
1999. {\it Proc. Natl. Acad. Sci. USA} 96:11311-11316

\item
Plaxco, K. W., K. T. Simons, and D. Baker.  
Contact order, transition state placement and the refolding rates
of single domain proteins.  1998. {\it J. Mol. Biol.} 277:985-994

\item
Plaxco, K. W., K. T. Simons, I. Ruczinski, D. Baker.  
Topology, stability, sequence, and length: defining the determinants 
of two-state protein folding kinetics.  2000. {\it Biochemistry} 39:11177-11183

\item
Plotkin, S. S., and J. N. Onuchic.
Investigation of routes and funnels in protein folding by free energy
functional methods.
2000. {\it Proc. Natl. Acad. Sci. USA} 97:6509-6514

\item
Settanni, G., T. X. Hoang, C. Micheletti, and A. Maritan.  
Folding pathways of prion and doppel.  
2002. SISSA preprint pp. http://xxx.lanl.gov/abs/cond-mat/0201196 

\item
Takada, S., and P. G. Wolynes.  
Microscopic theory of critical folding nuclei and reconfiguration activation
barriers in folding proteins.  1997. {\it J. Chem. Phys.} 107:9585-9598

\item
Takada, S.  Go-ing for the prediction of protein folding mechanism.  1999. 
{\it Proc. Natl. Acad. Sci. USA} 96:11698-11700.

\item
Thirumalai, D. 
>From minimal models to real proteins: time scales for protein folding.  
1995. {\it J. Physique I} 5:1457-1467

\item
Tsai, J., R. Taylor, C. Chothia, and M. Gerstein.  
The packing density in proteins: Standard radii and volumes.  
1999. {\it J. Mol. Biol.}  290:253-266

\item
Unger, R., and J. Moult.  
Local interactions dominate folding in a simple protein model.  1996. 
{\it J. Mol. Biol.} 259:988-994

\item
Veitshans, T., D. Klimov, and D. Thirumalai.  
Protein folding kinetics: time scales, pathways and energy landscapes
in terms of sequence-dependent properties.  1997. {\it Folding Des.} 2:1-22 

\item
Wolynes, P. G.  
Folding funnels and energy landscapes of larger proteins within the
capillarity approximation.  1997. {\it Proc. Natl. Acad. Sci. USA} 94:6170-6175

\item
Zhdanov, V. P.  
Folding time of ideal $\beta$ sheets vs. chain length.  1998. 
{\it Europhys Lett.} 42:577-581

\end{description}

\newpage

\begin{center}
{\bf FIGURE CAPTIONS}
\end{center}

Figure 1. Experimentally determined folding times based on tables
compiled by Plaxco et al. (Plaxco et al., 2000).
The solid circles, open hexagons, and stars are for the $\alpha$--$\beta$-,
$\alpha$-, and $\beta$- proteins respectively.\\

Figure 2. Experimentally determined folding times as split
into three structural classes.  The panels on the left hand side show
the dependence on $N$ whereas the panels on the right hand side
show the dependence on the relative contact order parameter.
Note that the time window in the middle
panels are shifted by two orders of magnitude compared to the
other panels. The {\em top} panels
corresponds to the following $\alpha$--$\beta$-proteins:
1div(56), 1gb1(56), 2ptl(63), 2ci2(65), 1aye(71), 1ubq(76), 1hdn(85),
2u1a(88), 1aps(98), and 2vik(126),
where the number in brackets indicates the value
of $N$ (in the case of 2ci2 there are 19 more amino acids but their
structure is undetermined).
The {\em middle} panels correspond to the following $\alpha$-proteins:
2pdd(43), 2abd(86), 1imq(86), 1lmb(92), 1hrc(104), 256b(106), and 1f63(154).
The {\em bottom} panels correspond to the following $\beta$-proteins:
1efn(57), 1csp(67), 1ten(89), and 1tit(89).
The papers by Plaxco et al. (Plaxco et al., 1998; Plaxco et al., 2000) also
contain data on several other proteins that are not shown here --
we restrict ourselves only to the proteins that we study through
simulations. (We had difficulties with the identification
of the proper structure files for the remaining proteins).
The subscript P in CO$_P$ signifies the criterion
of Plaxco et al. (Plaxco et al., 1998; Plaxco et al., 2000) for a
formation of a contact: two residues are considered to be contacting if
they contain non-hydrogen atoms within the distance of $6\AA$.
The symbols are as described in the caption of Figure 1.\\

Figure 3. The distribution of the effective contact lengths in T4 lysozyme
as determined by the procedure which is based on the van der Waals radii
of the atoms. The shaded region corresponds to the contacts that would
not be included if the cutoff of $7.5 \AA$ was adopted.\\

Figure 4. The distribution of the chirality parameter $C$ in 21
$\alpha$--$\beta$- proteins studied.

Figure 5. Times to establish contacts of a given sequence separation,
$|i-j|$ for crambin and for the indicated values of the
damping constant $\gamma$. The times are rescaled so that
$k$ is equal to 1, 6, and 12 for $\gamma$ equal to 2, 12 and 24 $m/\tau$ 
respectively and shown top to bottom. The symbols corresponding to $\gamma =12 m/\tau$
are reduced in size for clarity. The magnitude of the remaining symbols
indicates the size of the error bars.
The model used here corresponds to the Lennard-Jones contacts and the
contacts are determined based on the van der Walls radii.
The criterion for establishing a contact (for the first time)
is based on whether the two beads come within a distance of 1.5$\sigma _{ij}$
of each other. This figure illustrates existence of second order
effects in the dependence on $\gamma$ because the rescaling by $k$
brings the data points for a given event together but there is
no strict overlapping.\\

Figure 6.  The dependence of the folding time on temperature for
"synthetic" secondary structures of 16 monomers. The top two panels
are for the $\alpha$-helix system H16 and the bottom panel is for
the $\beta$-hairpin B16.
The dotted lines correspond to the chirality potential (with $\kappa$=1)
included and the solid lines are for the case when it is not.
In all cases, $R_c=7.5\AA$.
The top panel corresponds to the contact based criterion
whereas the other panels is for the criterion based on the cutoff
RMSD of $0.2\AA$.\\

Figure 7. The dependence of the folding time on temperature for
various Go models of crambin. The top panel is for the contact
cutoff range of $7.5\AA$ whereas the bottom panel is for
the locally calculated contact ranges.
On the top panel, the dotted line corresponds to the case with the
chirality potential and the solid line -- without.
On the bottom panel, both curves include the chirality potential.
Here, the solid (dashed) line is for the 6--12 (10--12) contact potential.
The arrows indicate values of the folding temperature $T_f$. The
heavier (lighter) arrow is for the 6--12 (10--12) potential.\\

Figure 8. The dependence of the folding time on temperature
for models of the protein 1tit. The symbols are as in Figure 6:
the thin solid line and the triangular data points are for $R_c=7.5\AA$
and no chirality; the dotted line with the square data points are
for $R_c=7.5\AA$ and with the chirality;
the thick solid line with the solid circular data points
are for the locally calculated $R_c$ and the Lennard-Jones contact
potential with the chirality; the dashed line with the open circular
data points are for the similar case with the 10--12 potential.
The arrows indicate the values of $T_f$
for the contacts of variable range: thick for the Lennard-Jones
case and thin for the 10--12 case.\\

Figure 9. The scaling of $t_{fold}$ with $N$ for the 51 proteins as
modeled by the 6--12 contact potential with the variable contact range.
The data are split into the $\alpha$--$\beta$-, $\alpha$-, and
$\beta$-proteins as indicated.  The lines indicate the power
law behavior with the $\lambda$ exponent displayed in the right hand corner
of each panel. The error bars in the exponent are of order
$\pm 0.2$. The folding times are calculated at $T_{min}$.
The correlation levels of the points shown are 81\%, 97\% and 94\% for 
the top, middle and bottom panels respectively. \\

Figure 10. This Figure replots the data points of Figure 8
in one panel.
For clarity, two of the most distant outlayers in each
class are not shown. The solid, dotted, and broken lines correspond
to the slopes of 3.2, 2.5, and 1.7 respectively.
The correlation level is 87\%.\\

Figure 11. Same as Figure 8 but the folding times are determined at
$T_f$ instead at $T_{min}$.
The data points represented by the arrows indicate values which
are significantly off the frame of the figure (for which
only the lower bound of 30000 $\tau$ is known).
The correlation levels are 83\%, 88\% and 77\% for
the top to bottom panels respectively. \\

Figure 12. The role of the chirality potential on the folding times
for the $\alpha$-proteins. The hexagons are the data points
shown in the middle panel of Figure 8 whereas the crosses correspond
to the results obtained for $\kappa$=0.\\

Figure 13. The values of $T_{min}$ and $T_f$ shown vs. $N$
for the Lennard-Jones potential with the variable contact range.
The data points are divided into the three structural classes.\\

Figure 14. The dependence of $t_{fold}$, $T_{min}$, and $T_f$
on the relative contact order parameter for the Lennard-Jones contact
potential with the variable contact range. The data symbols
indicate the structural classes and are identical to those
used in Figures 8, 9, 10, and 11.\\

Figure 15. Same as in Figure 8 but for the 10--12 contact potential.
The correlation levels are 88\%, 98\% and 91\% from top to bottom. \\

Figure 16. Same as in Figure 12 but for the 10--12 contact potential.\\

Figure 17. Same as in Figure 8 but for the Lennard-Jones potential
with $R_c=7.5\AA$.
The correlation levels are 83\%, 91\% and
93\% for the top to bottom panels respectively.\\

Figure 18. Same as in Figure 9 but for the cutoff of $7.5 \AA$
in the range of the contact potential. The solid line has a slope of 2.5.
The correlation level for all of the points is 88\%.\\

Figure 19. Same as in Figure 12 but for $R_c=7.5\AA$.\\

Figure 20. Logarithm of the folding time vs. CO$\times N^{0.6}$
for the three structural classes. The data correspond to the
Lennard-Jones potential with the variable range.
The left hand panels are for $T=T_{min}$ and the right hand panels for
$T=T_f$. Note that the horizontal scale in this figure is linear,
not logarithmic as in most previous figures. The arrows, like in
Figure 11, indicate data points which are significantly off the
scale of the frame of the figure.\\

Figure 21. The data of figure 9 redisplayed on the log - linear 
plane. The dashed lines indicate fits to the exponential law
$t_{fold} \sim exp(b/\xi)$ with the values of $\xi$ shown in the
right hand corner of each panel.  
The correlation levels are 
75\%, 94\% and 95\%  for the top to the bottom panels respectively.
The overall correlation level
is 82\% whereas for the power law fit it is 86\%. The corresponding
numbers for the 10--12 potential and the Lennard-Jones with the cutoff 
of $7.5\AA$ are 87\%, 89\% and 81\%, 88\%. The fitted values of $\xi$
for the 10--12 potential are about the same as for the Lennard-Jones
case.\\

Abbreviations used: PDB, Protein Data Bank; NMR, nuclear magnetic resonance. 

\onecolumn

\newpage
\begin{figure}
\vspace*{-0.5cm}
\epsfxsize=3.8in
\centerline{\epsffile{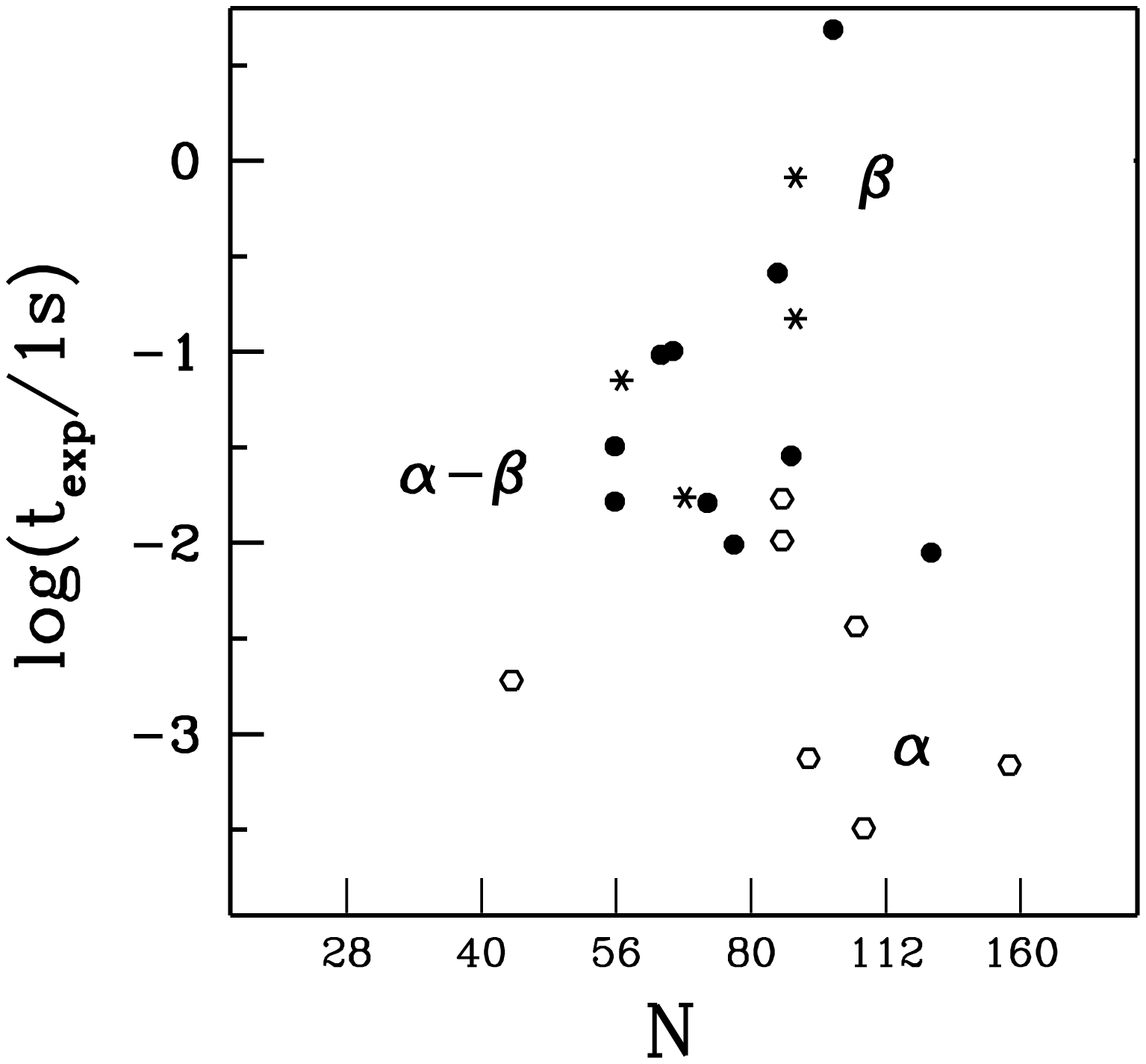}}
\vspace*{0.5cm}
\caption{ }
\end{figure}

\begin{figure}
\vspace*{-0.5cm}
\epsfxsize=3.8in
\centerline{\epsffile{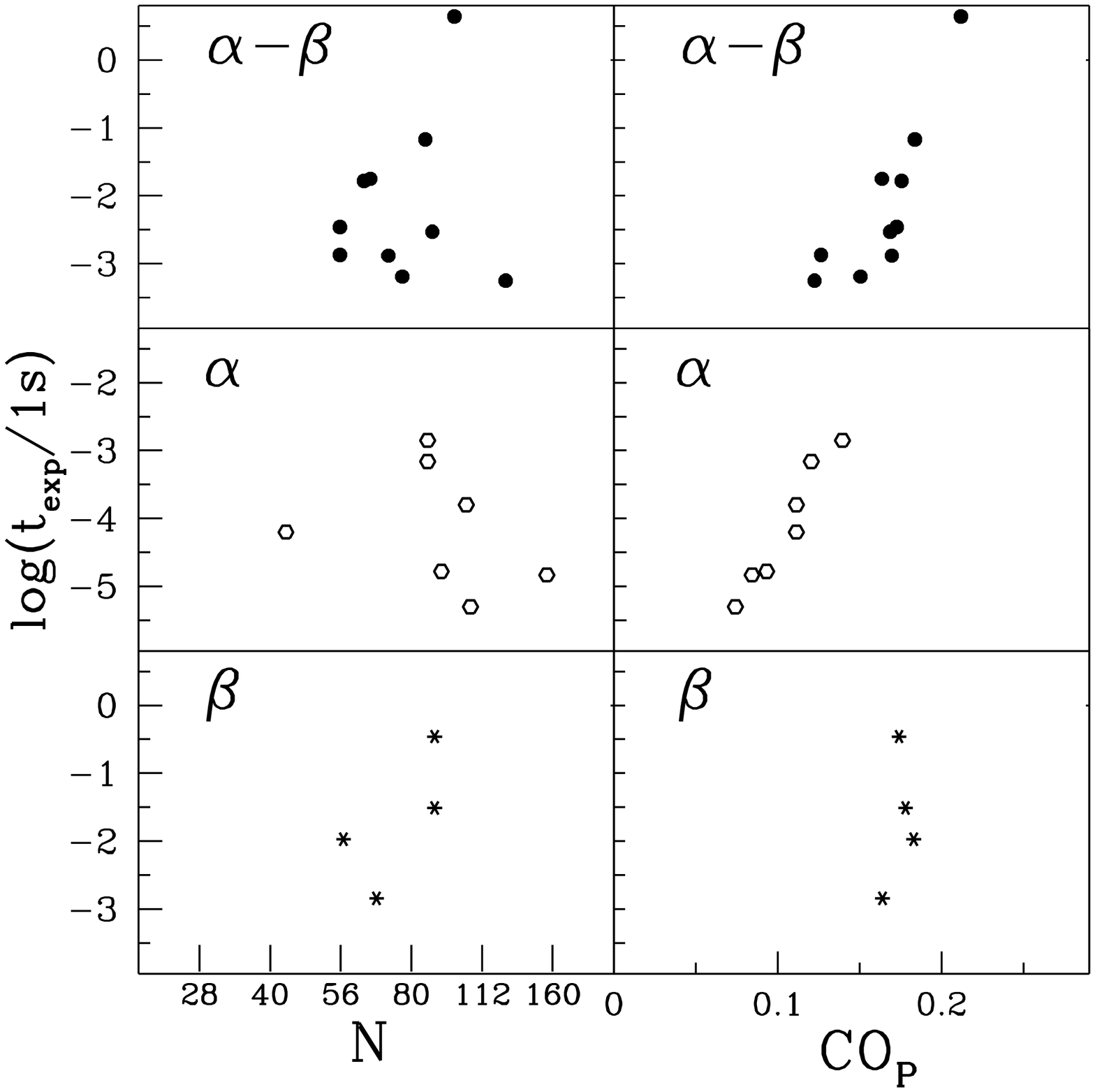}}
\vspace*{0.5cm}
\caption{ }
\end{figure}

\begin{figure}
\vspace*{-0.5cm}
\epsfxsize=3.8in
\centerline{\epsffile{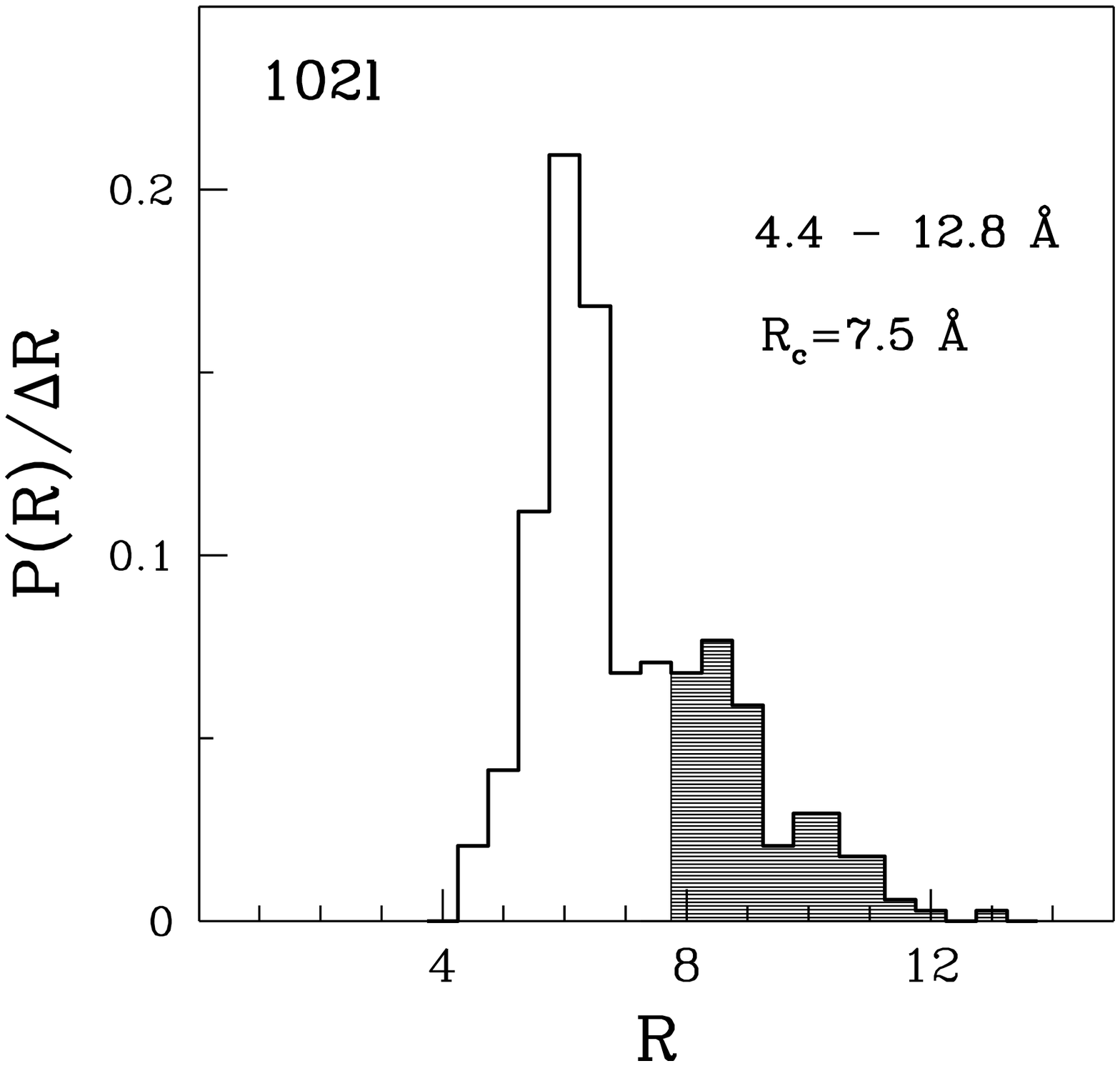}}
\vspace*{0.5cm}
\caption{ }
\end{figure}

\begin{figure}
\vspace*{-0.5cm}
\epsfxsize=3.8in
\centerline{\epsffile{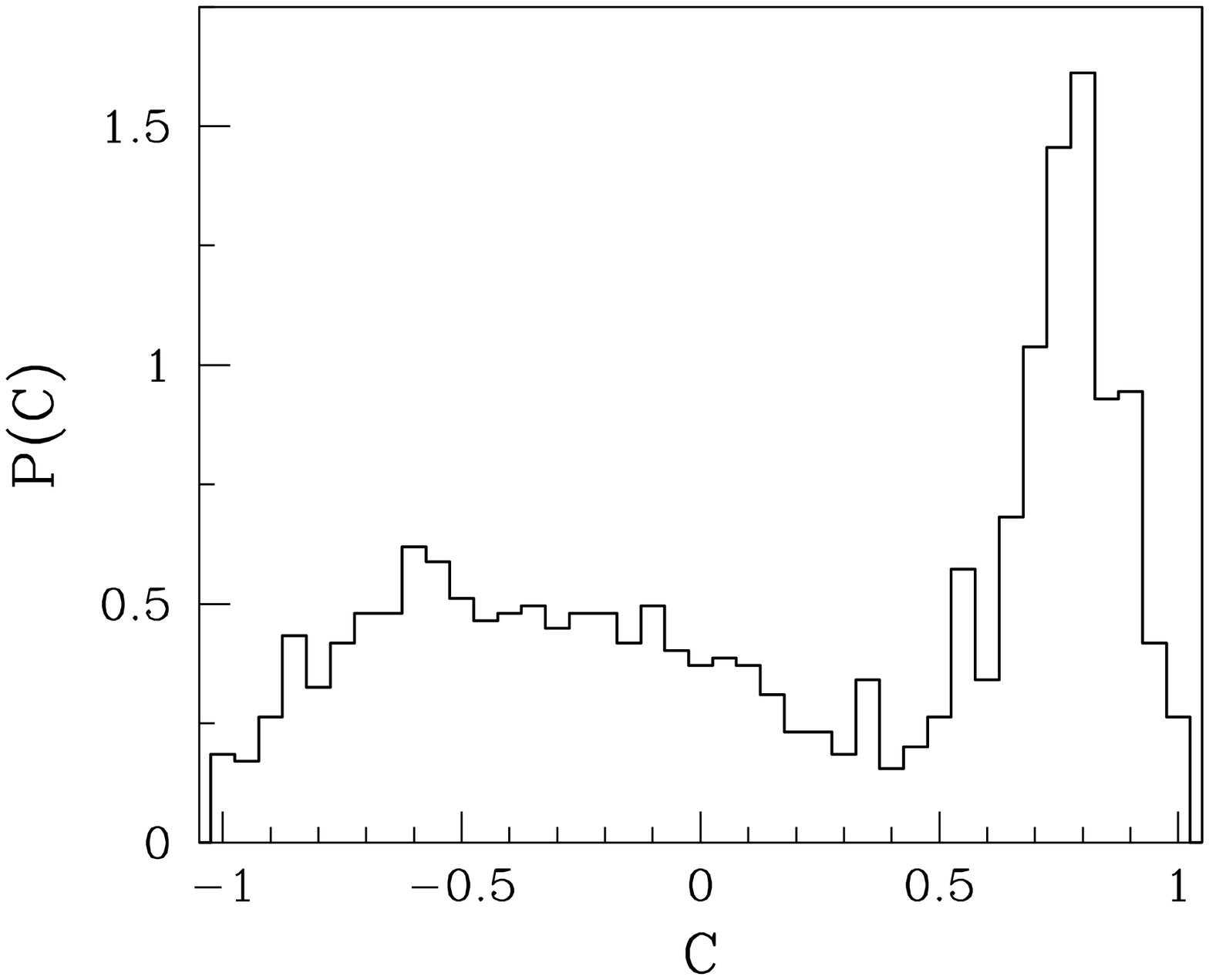}}
\vspace*{0.5cm}
\caption{ }
\end{figure}

\begin{figure}
\vspace*{-0.5cm}
\epsfxsize=3.8in
\centerline{\epsffile{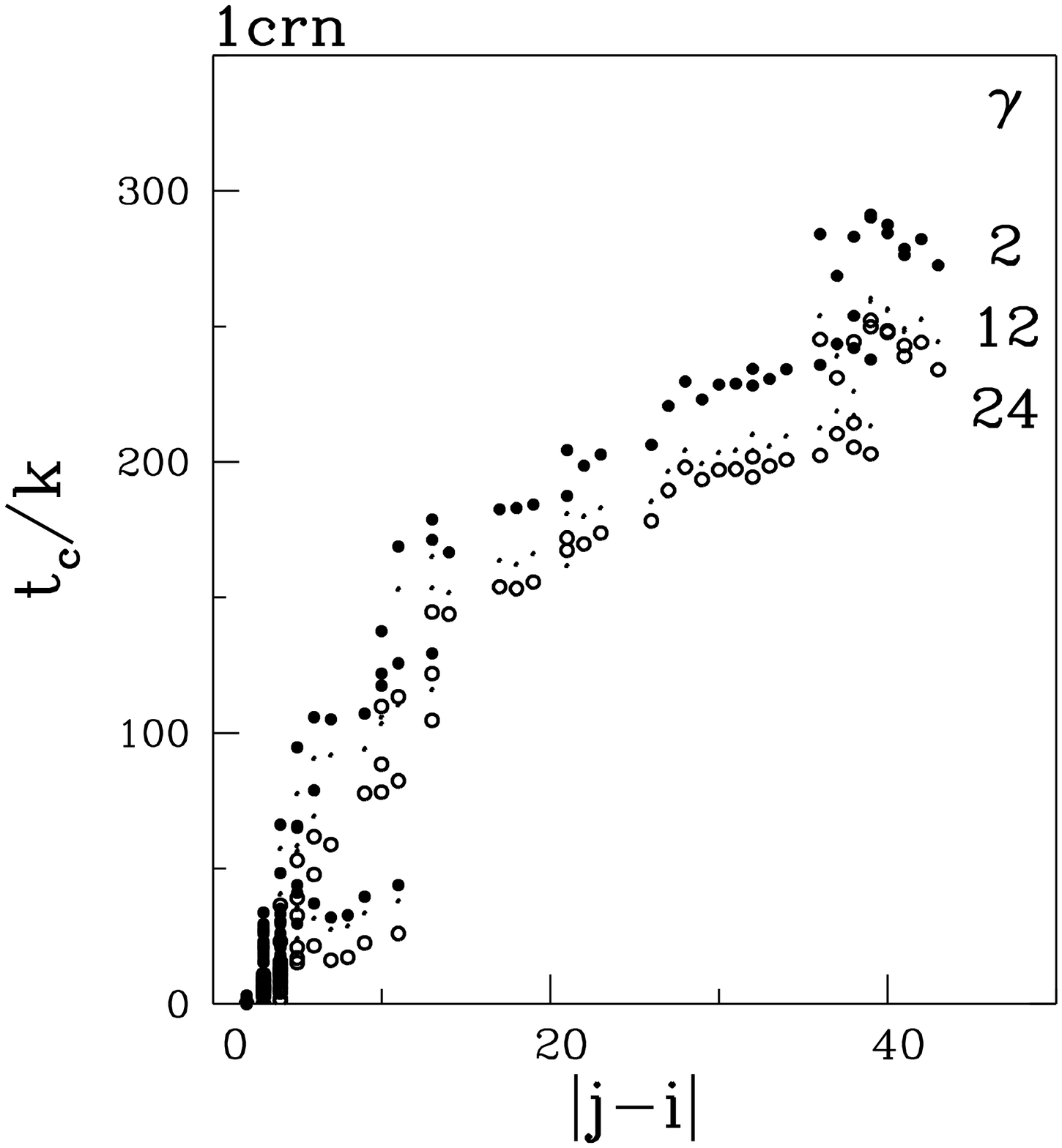}}
\vspace*{0.5cm}
\caption{ }
\end{figure}

\begin{figure}
\vspace*{-0.5cm}
\epsfxsize=3.8in
\centerline{\epsffile{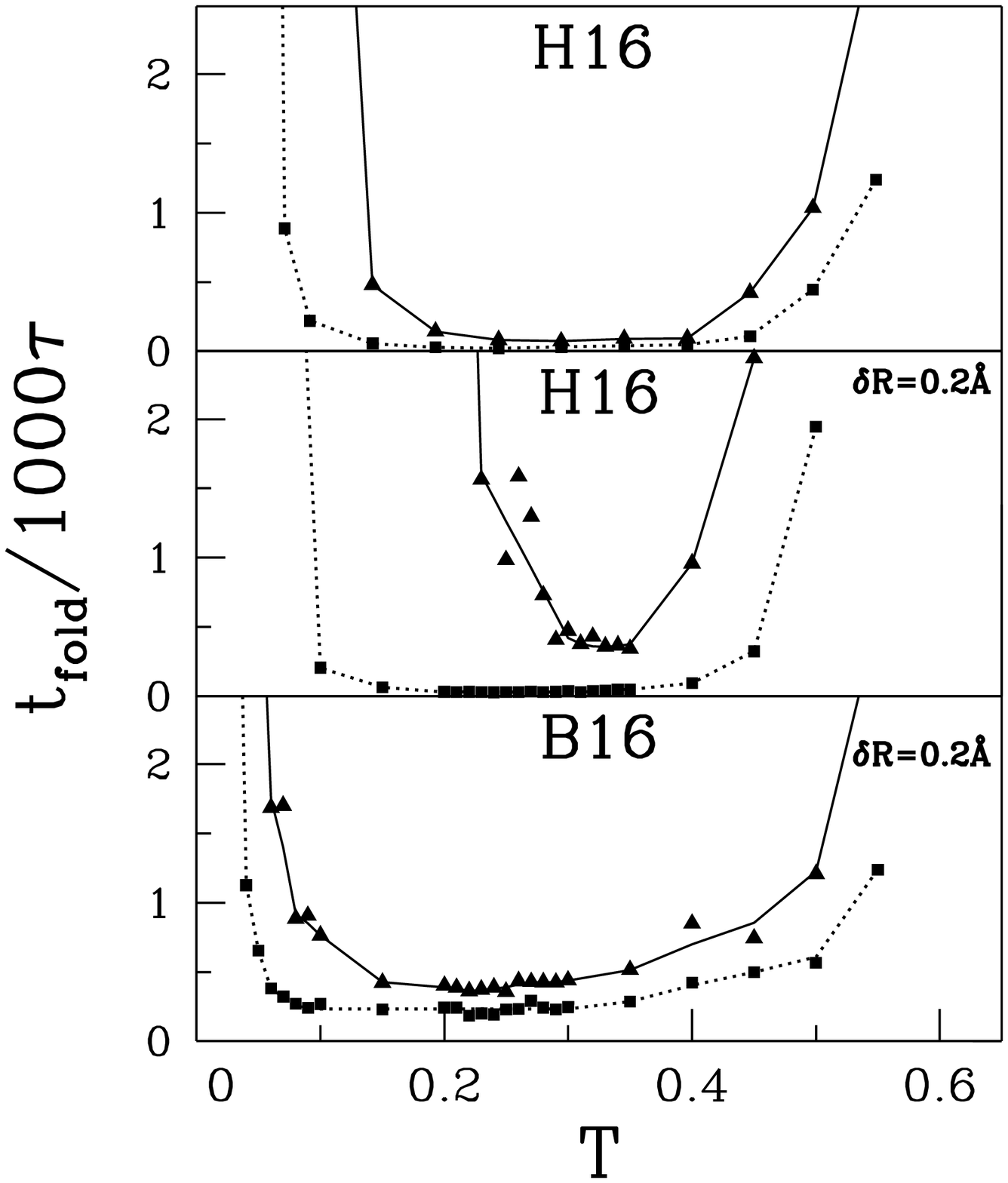}}
\vspace*{0.5cm}
\caption{ }
\end{figure}

\begin{figure}
\vspace*{-0.5cm}
\epsfxsize=3.8in
\centerline{\epsffile{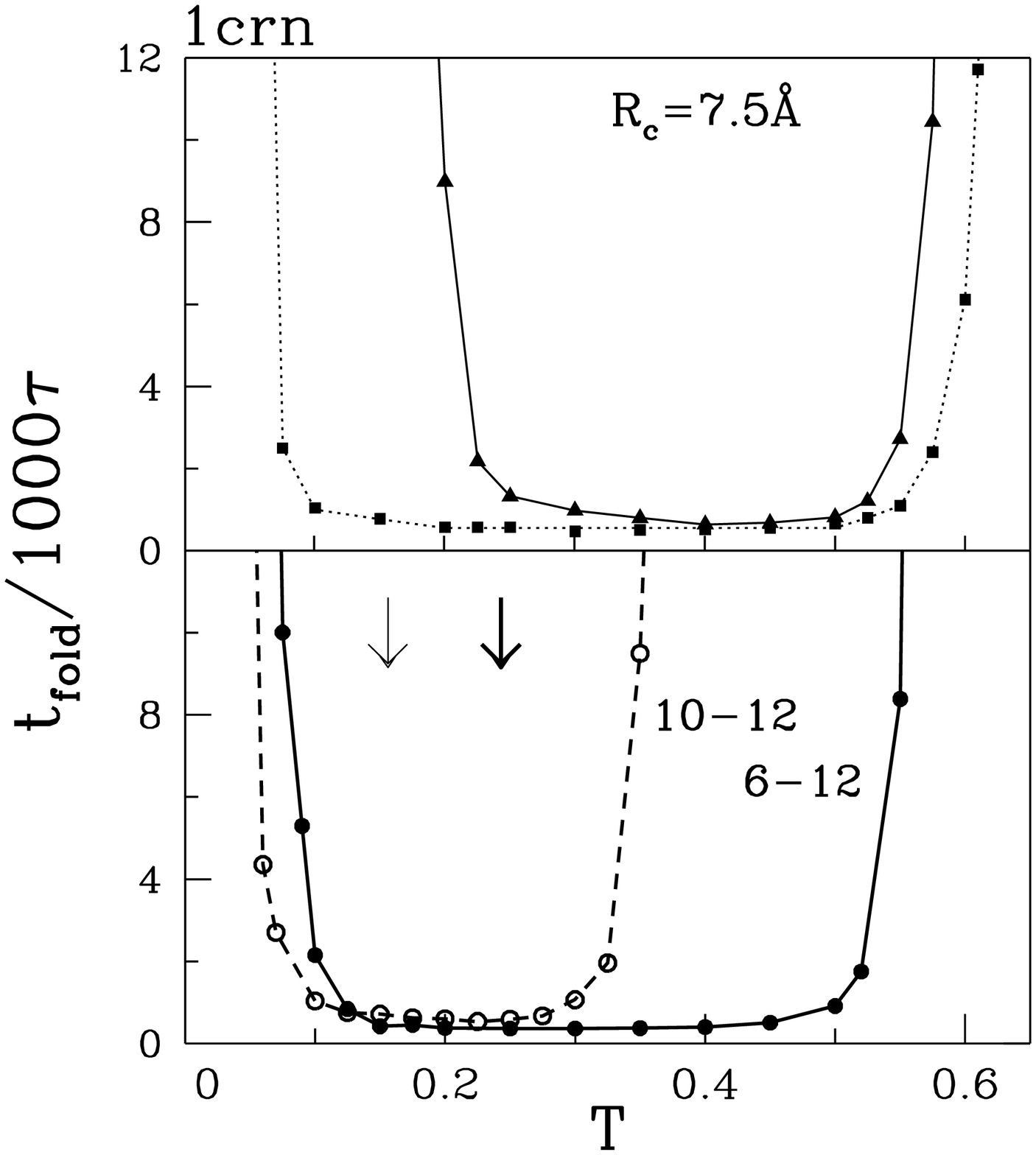}}
\vspace*{0.5cm}
\caption{ }
\end{figure}

\begin{figure}
\vspace*{-0.5cm}
\epsfxsize=3.8in
\centerline{\epsffile{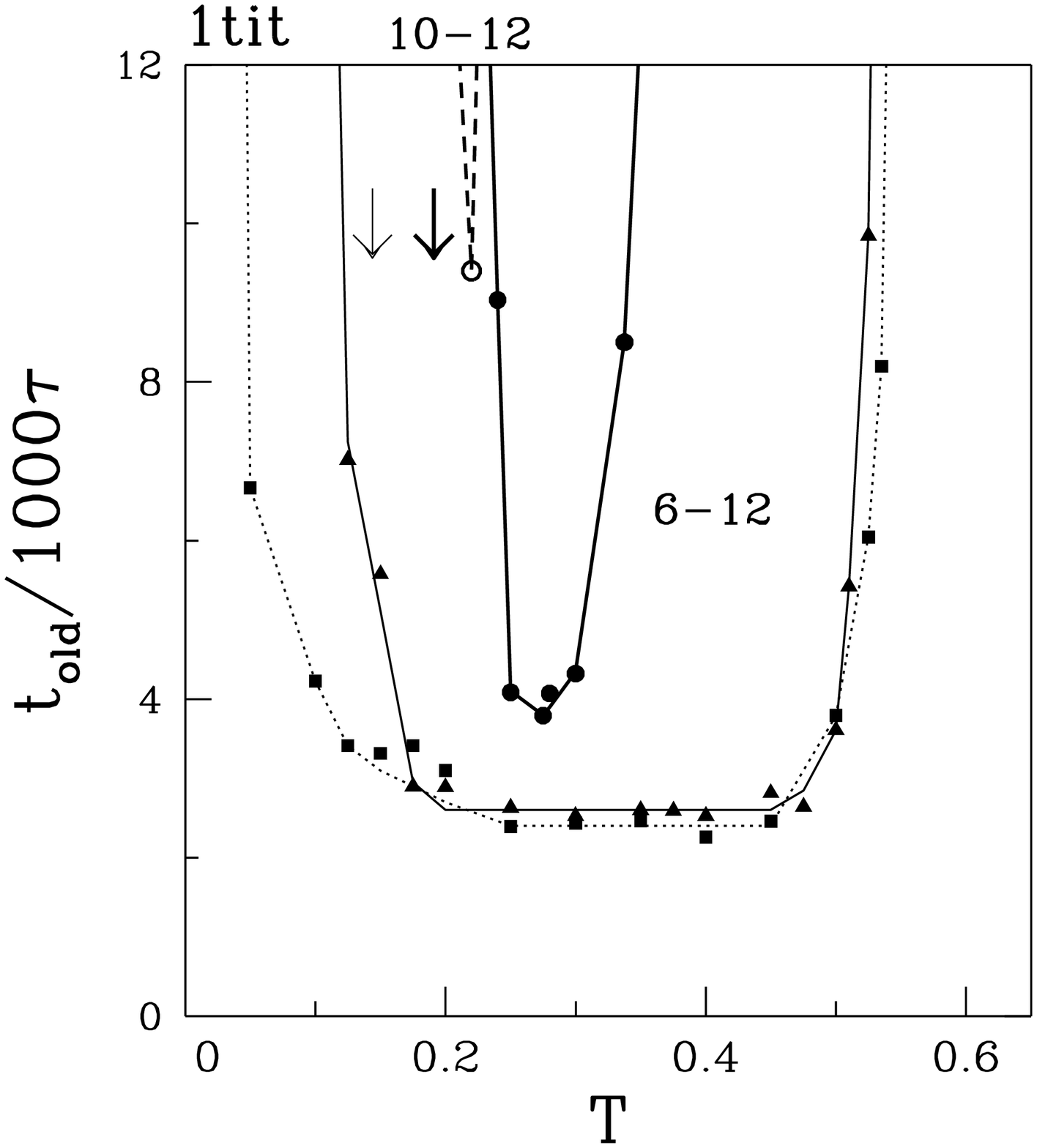}}
\vspace*{0.5cm}
\caption{ }
\end{figure}

\begin{figure}
\vspace*{-0.5cm}
\epsfxsize=3.8in
\centerline{\epsffile{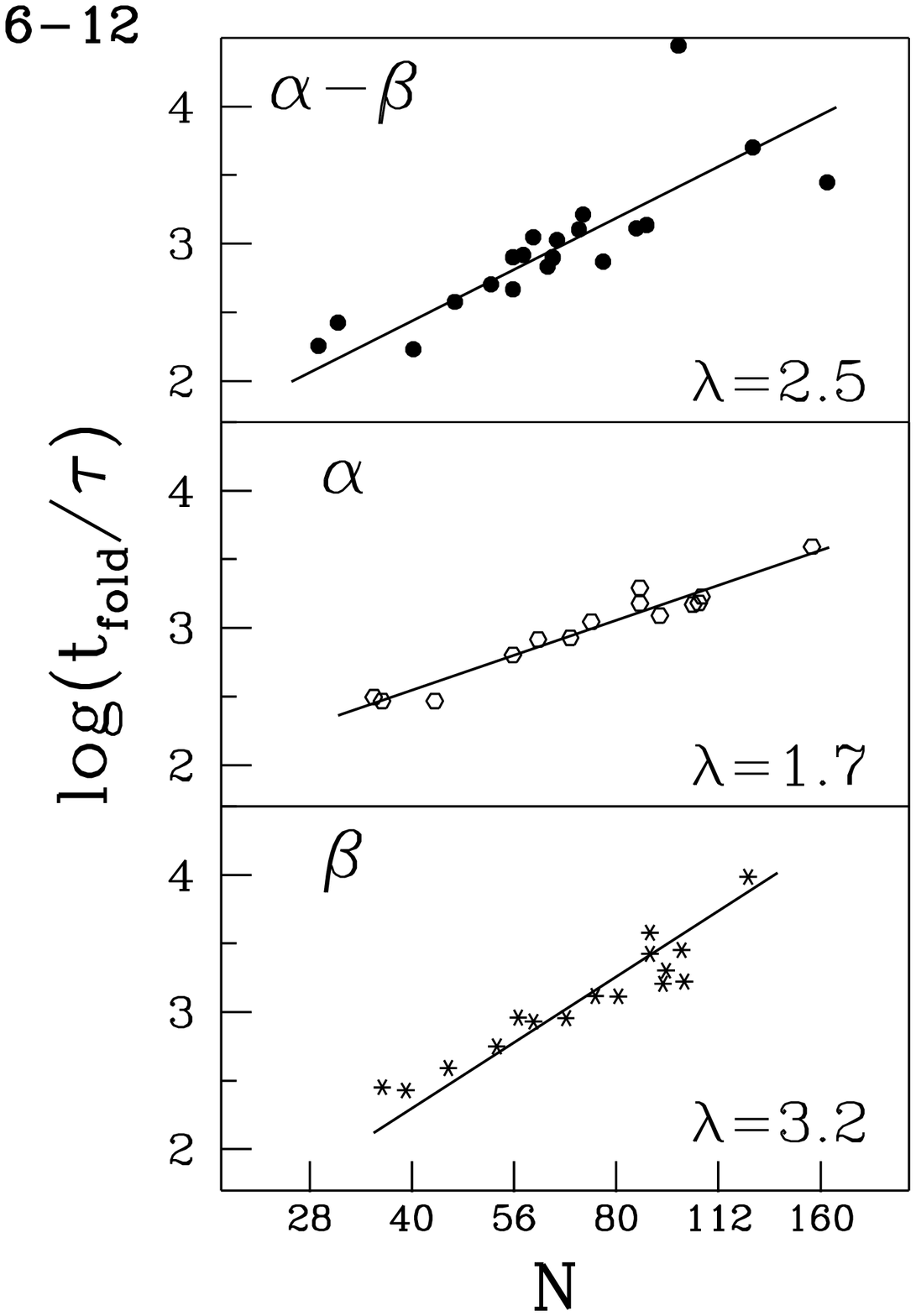}}
\vspace*{0.5cm}
\caption{ }
\end{figure}

\begin{figure}
\vspace*{-0.5cm}
\epsfxsize=3.8in
\centerline{\epsffile{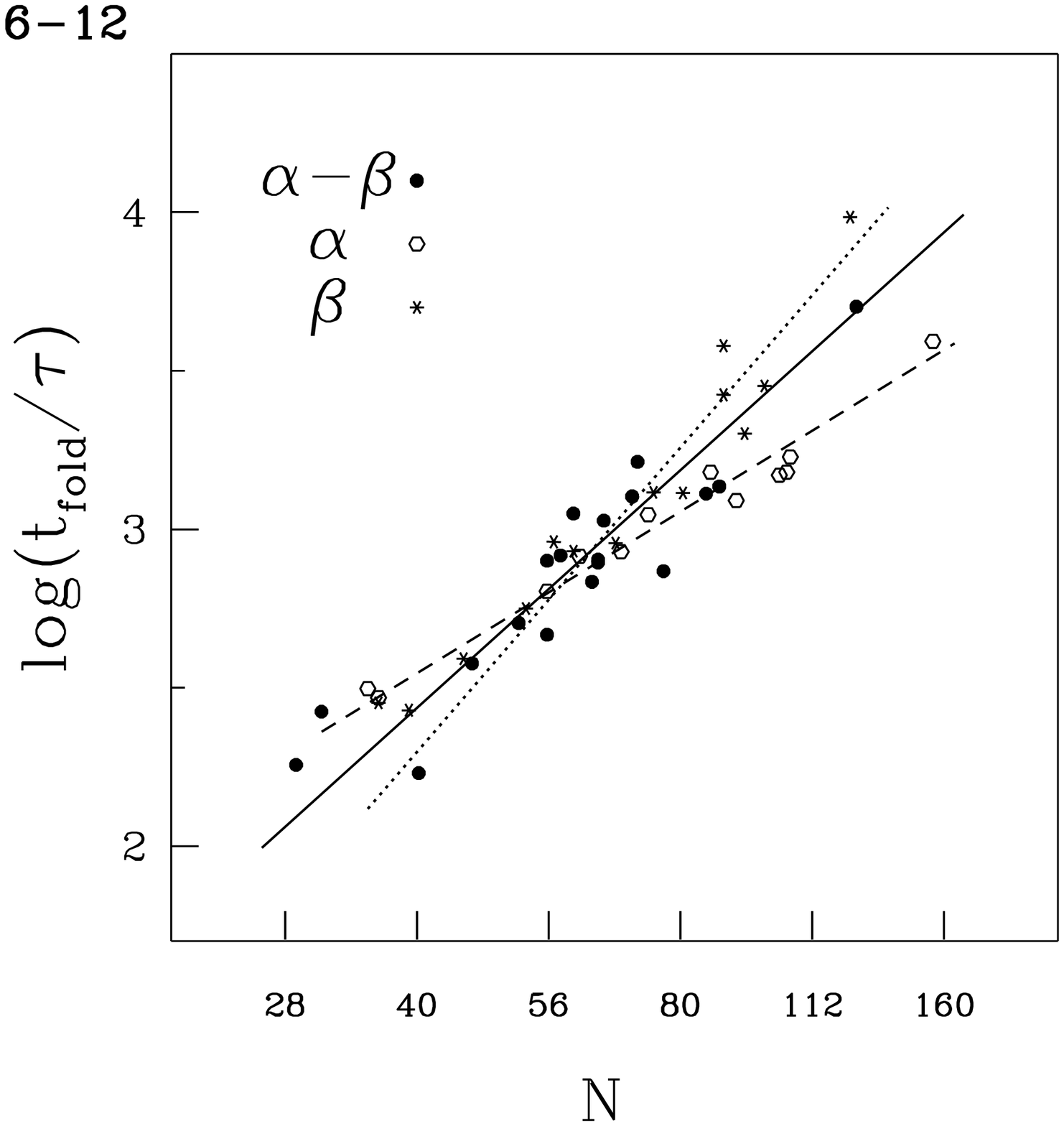}}
\vspace*{0.5cm}
\caption{ }
\end{figure}

\begin{figure}
\vspace*{-0.5cm}
\epsfxsize=3.8in
\centerline{\epsffile{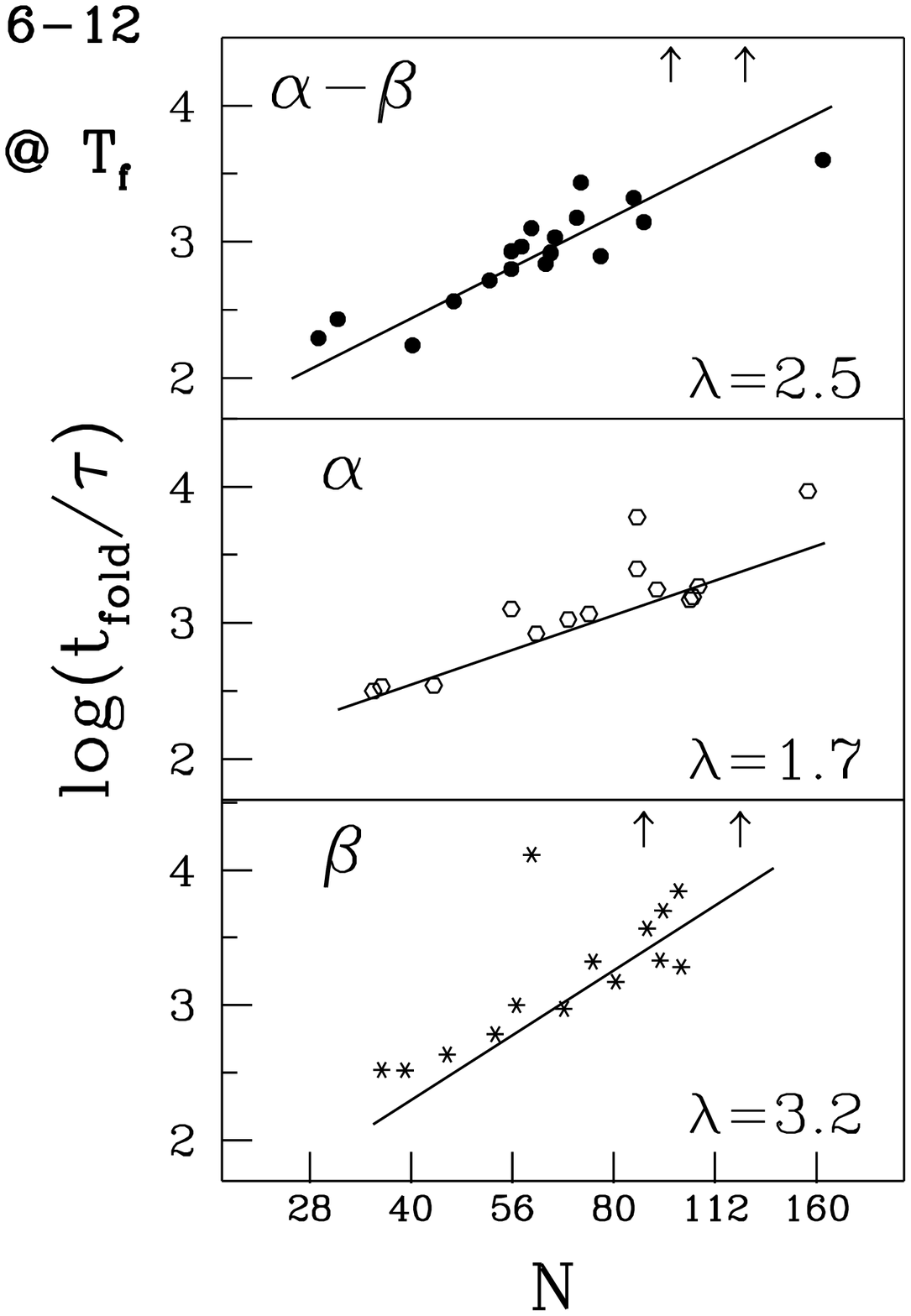}}
\vspace*{0.5cm}
\caption{ }
\end{figure}

\begin{figure}
\vspace*{-0.5cm}
\epsfxsize=3.8in
\centerline{\epsffile{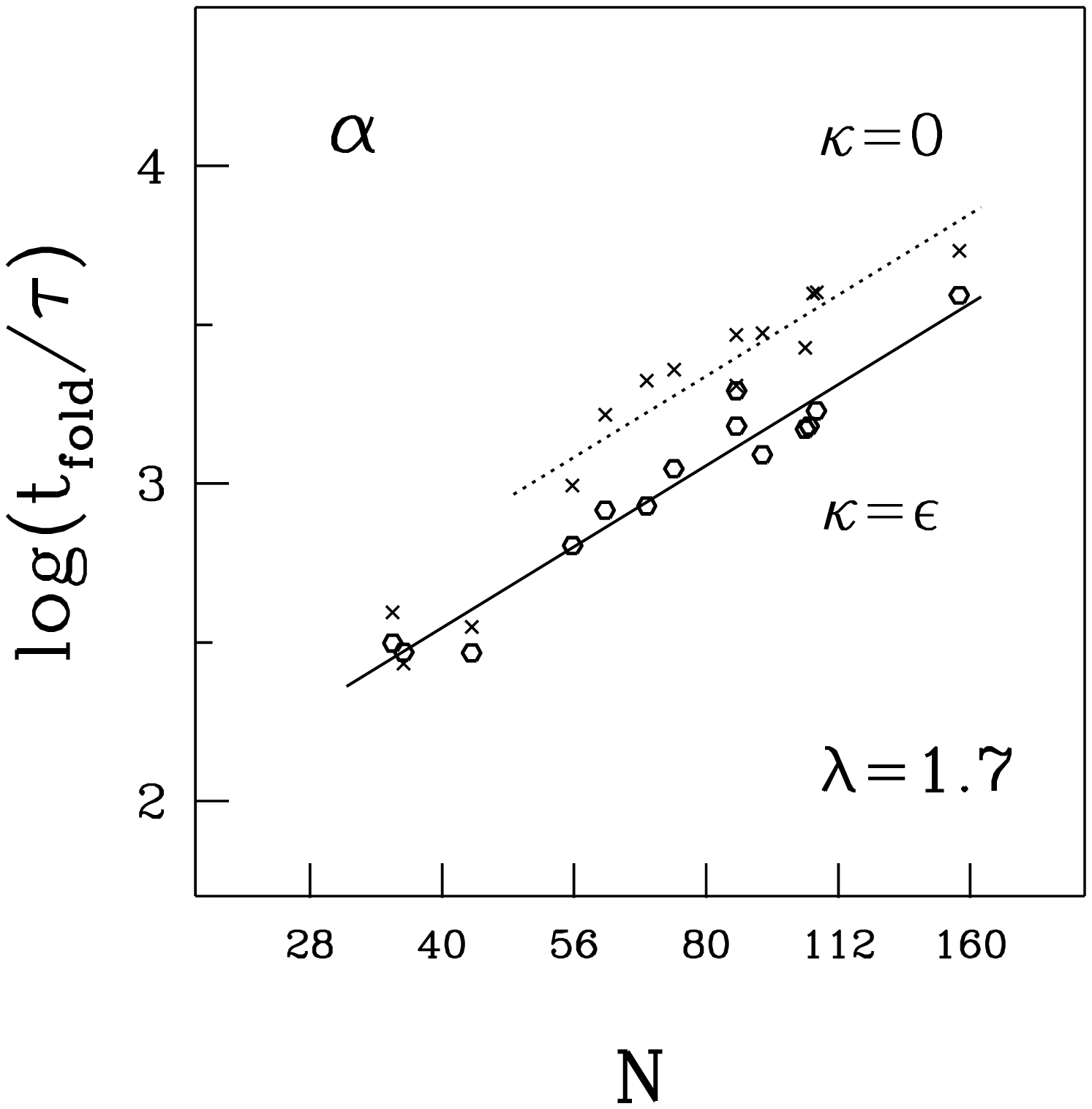}}
\vspace*{0.5cm}
\caption{ }
\end{figure}

\begin{figure}
\vspace*{-0.5cm}
\epsfxsize=3.8in
\centerline{\epsffile{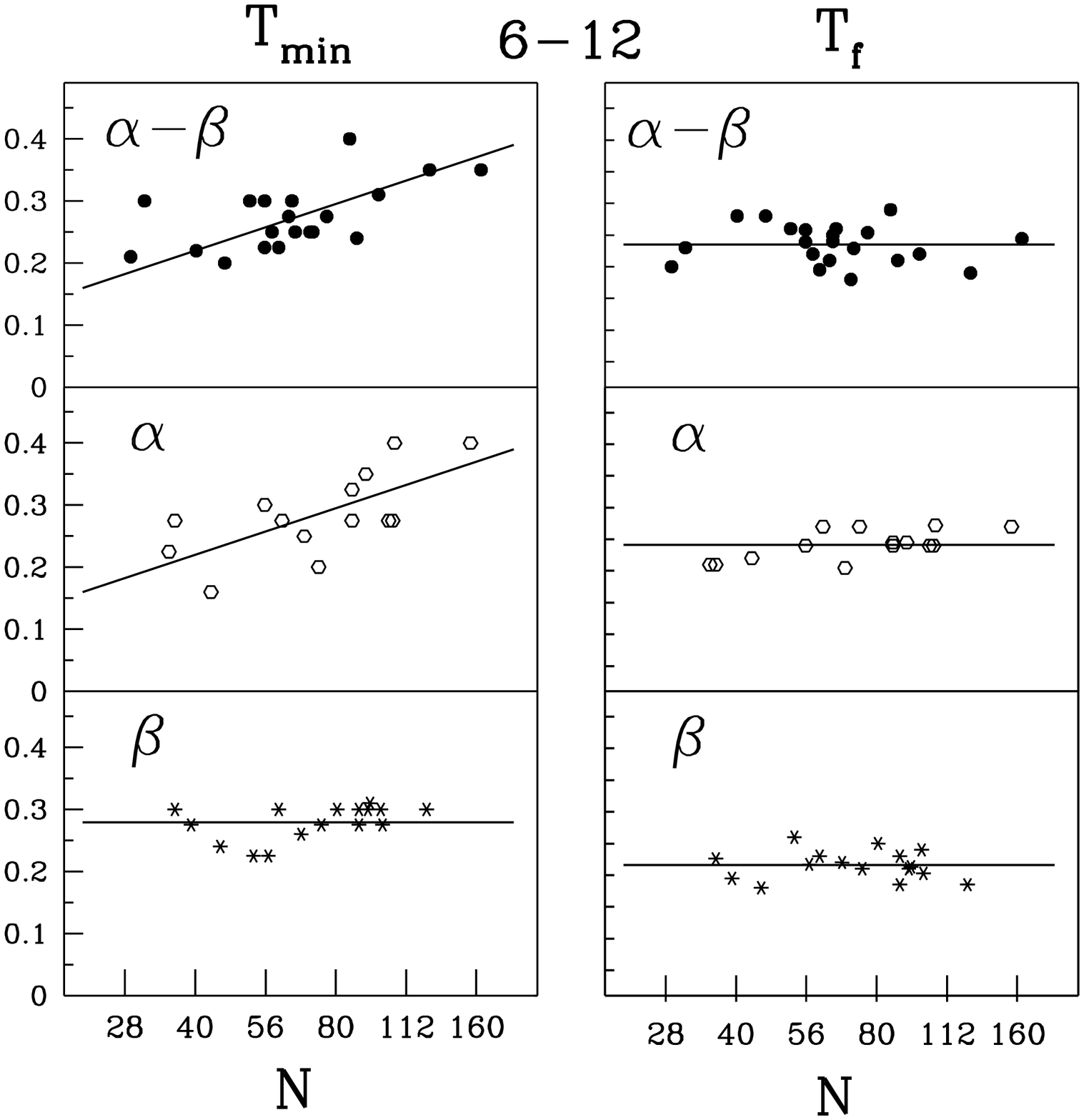}}
\vspace*{0.5cm}
\caption{ }
\end{figure}

\begin{figure}
\epsfxsize=3.8in
\centerline{\epsffile{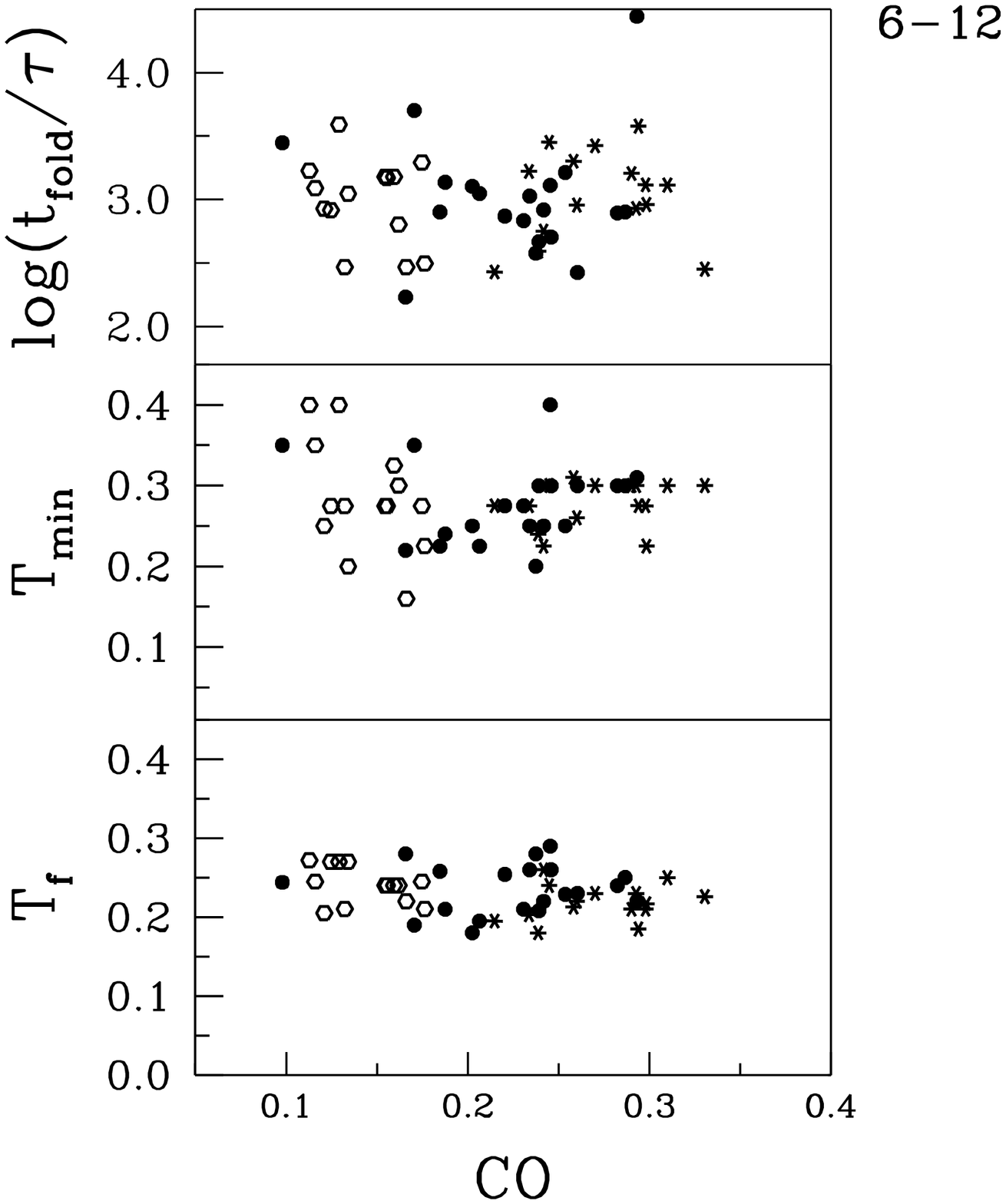}}
\vspace*{0.5cm}
\caption{ }
\end{figure}

\begin{figure}
\vspace*{-0.5cm}
\epsfxsize=3.8in
\centerline{\epsffile{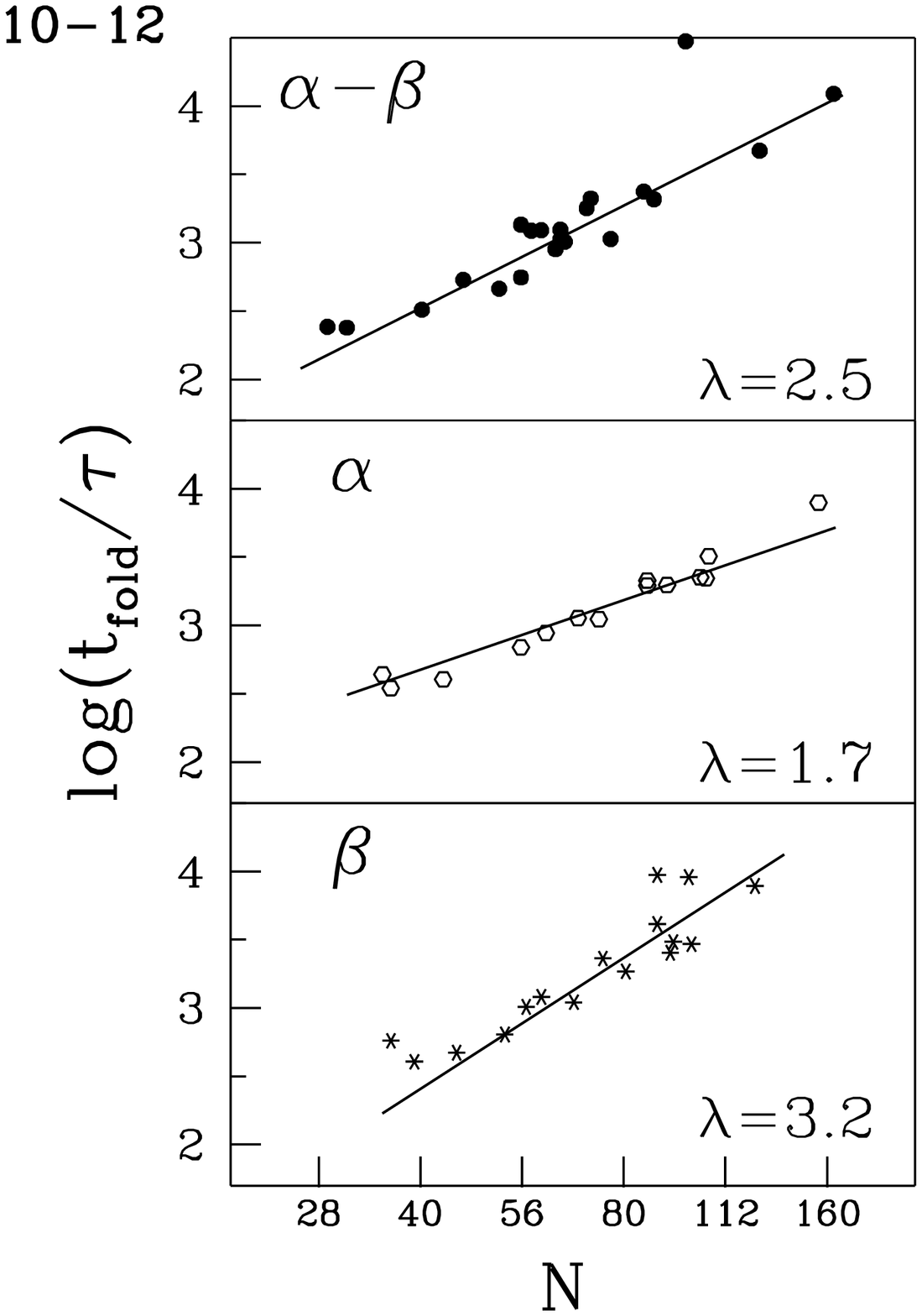}}
\vspace*{0.5cm}
\caption{ }
\end{figure}


\begin{figure}
\vspace*{-0.5cm}
\epsfxsize=3.8in
\centerline{\epsffile{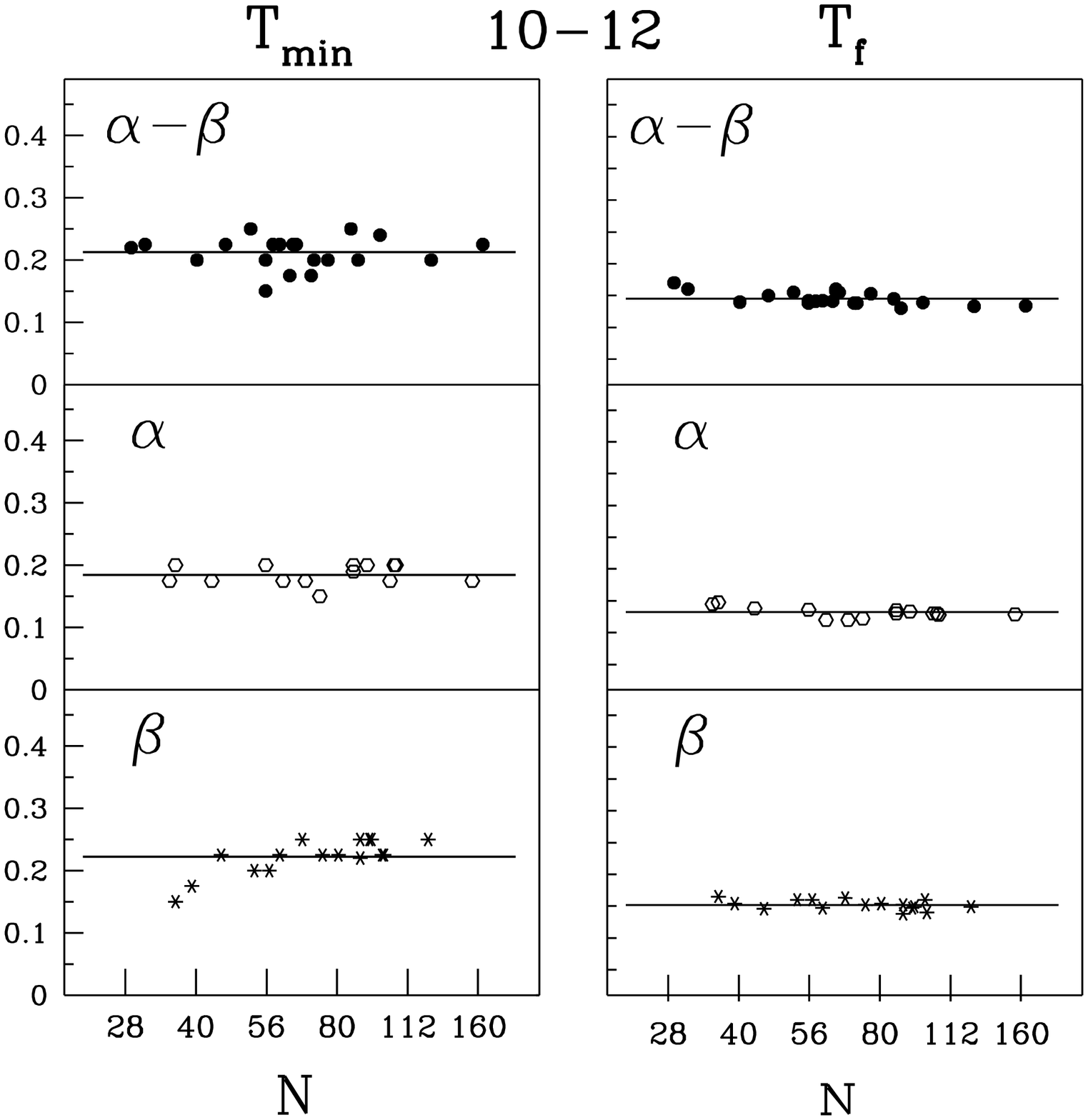}}
\vspace*{0.5cm}
\caption{ }
\end{figure}


\begin{figure}
\vspace*{-0.5cm}
\epsfxsize=3.8in
\centerline{\epsffile{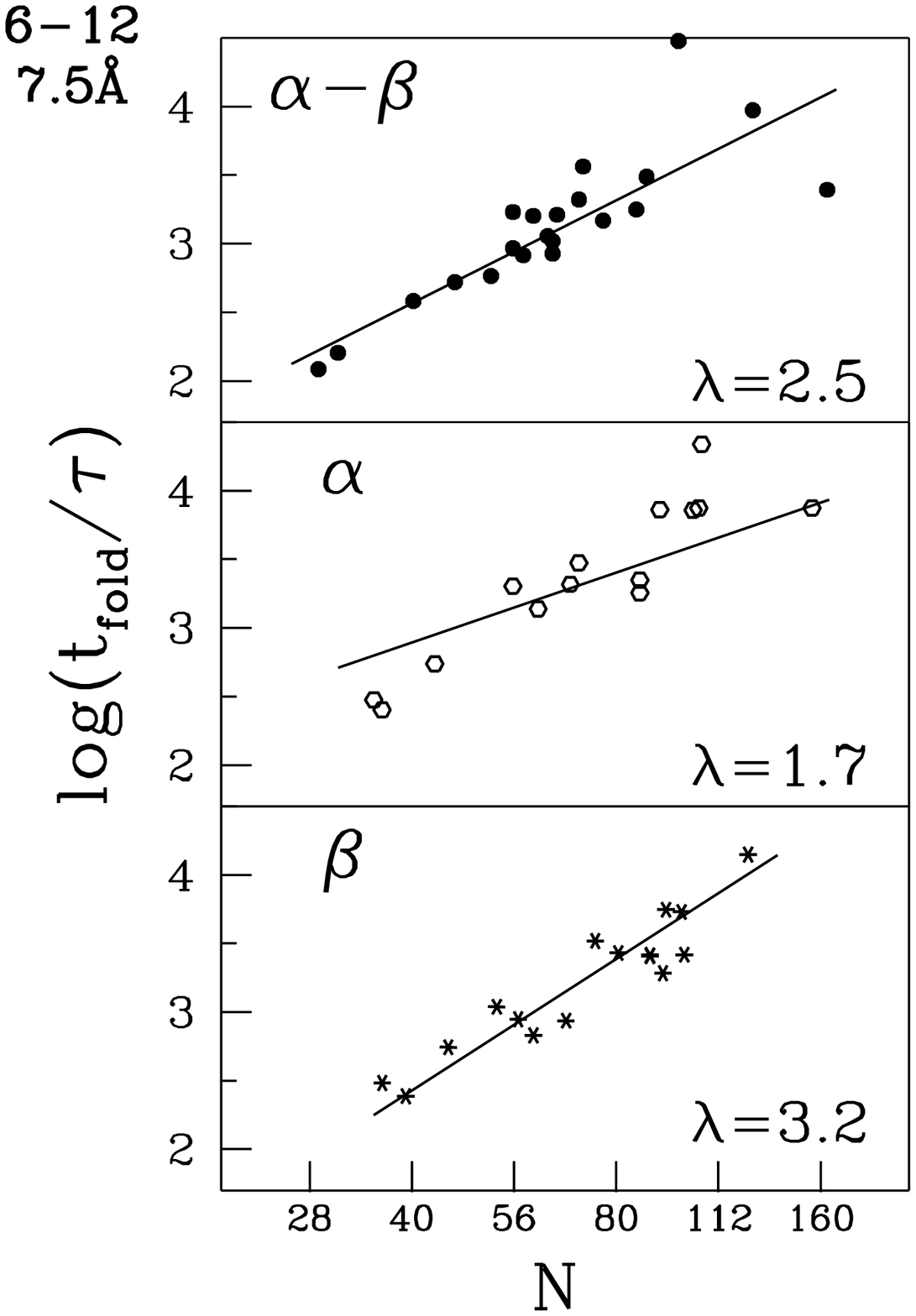}}
\vspace*{0.5cm}
\caption{ }
\end{figure}

\begin{figure}
\vspace*{-0.5cm}
\epsfxsize=3.8in
\centerline{\epsffile{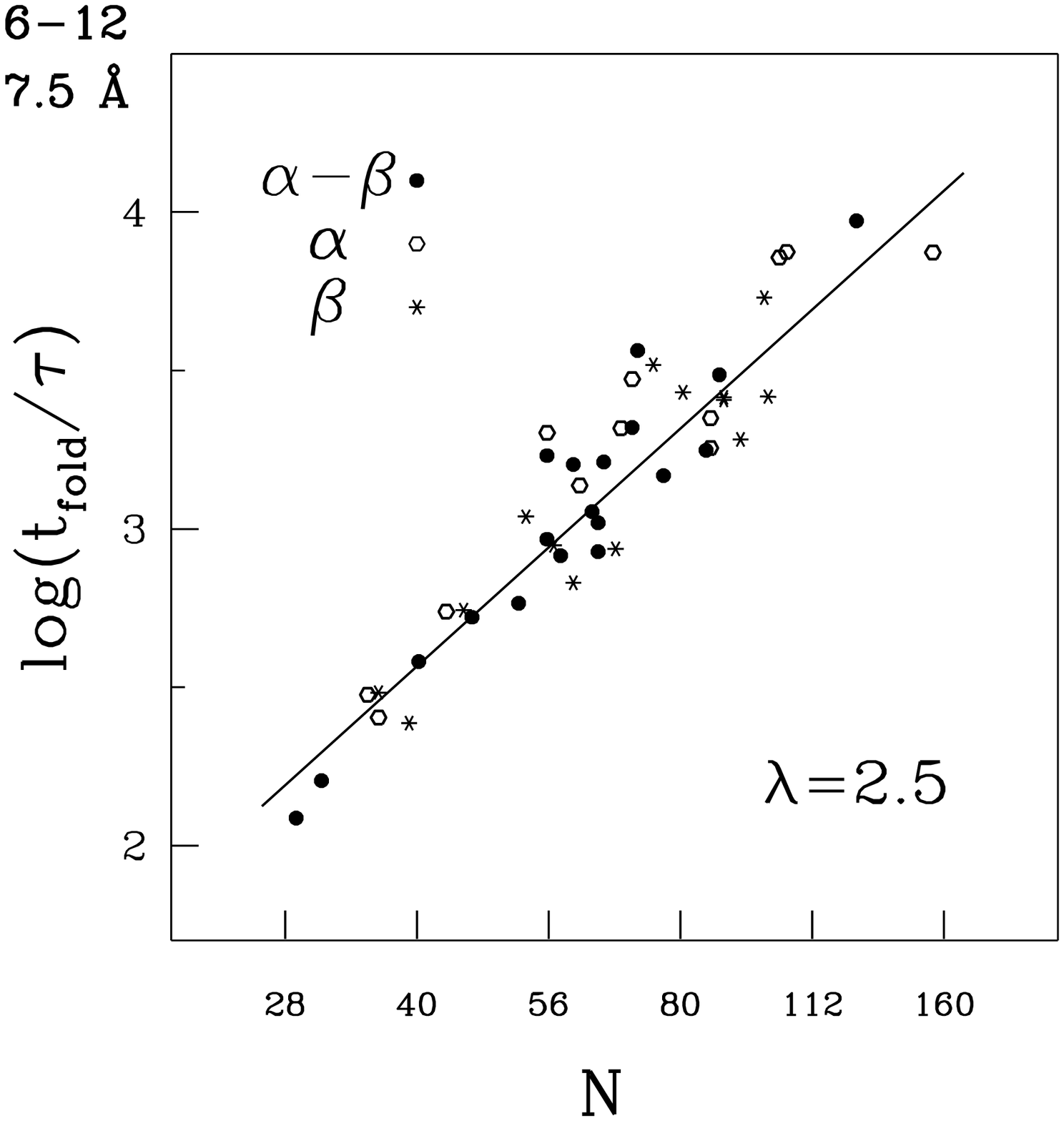}}
\vspace*{0.5cm}
\caption{ }
\end{figure}

\begin{figure}
\vspace*{-0.5cm}
\epsfxsize=3.8in
\centerline{\epsffile{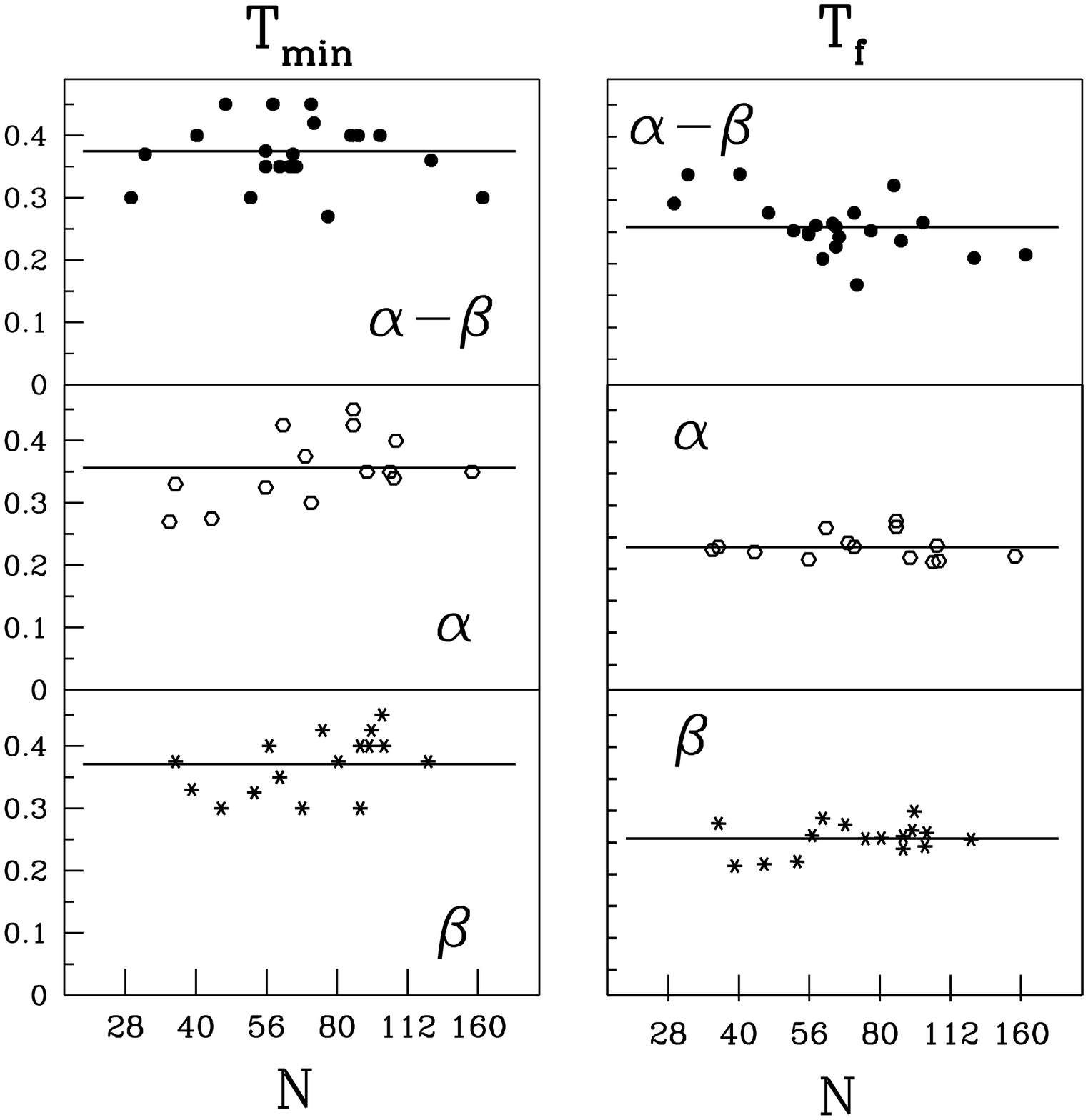}}
\vspace*{0.5cm}
\caption{ }
\end{figure}

\begin{figure}
\vspace*{-0.5cm}
\epsfxsize=3.8in
\centerline{\epsffile{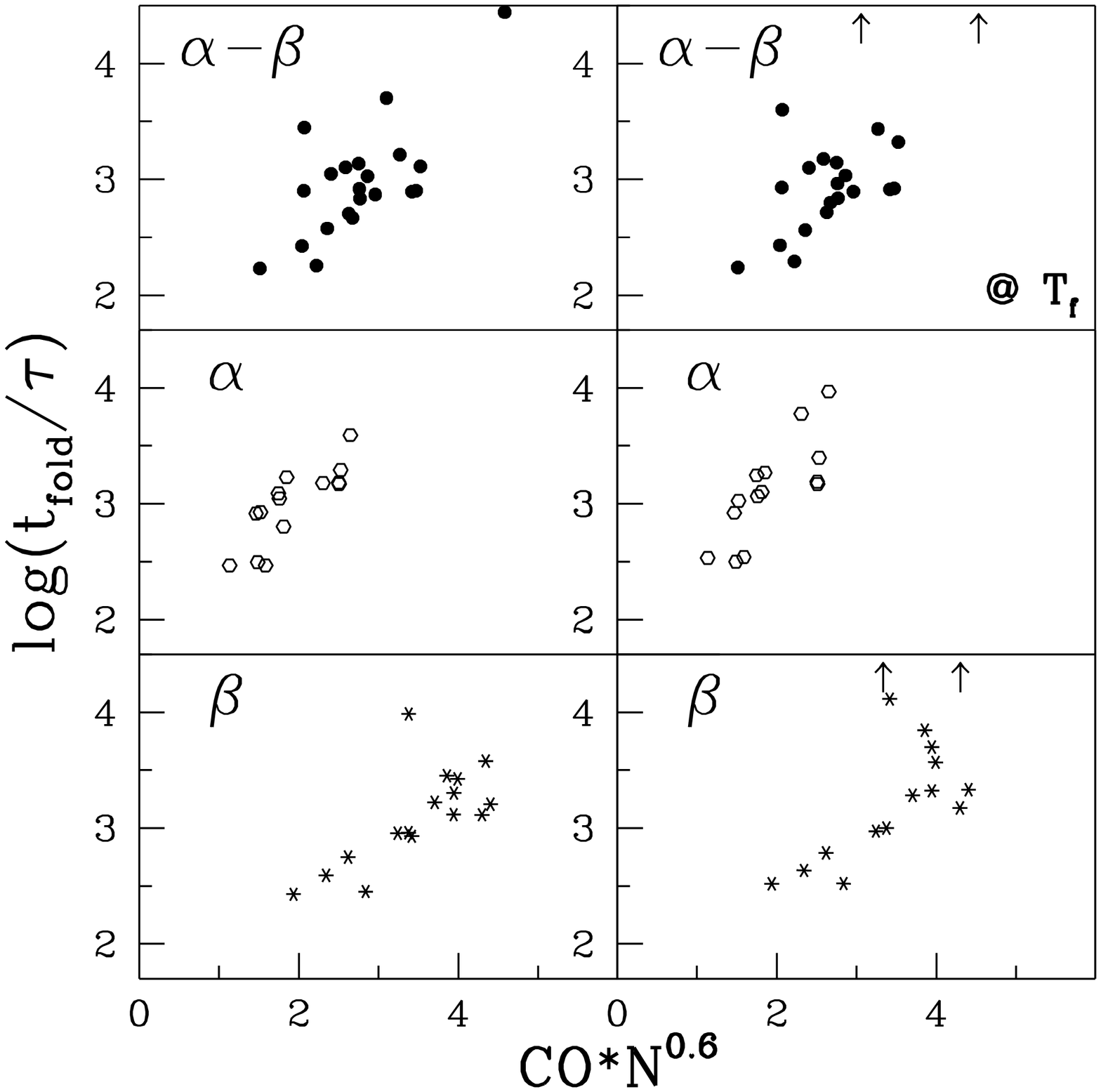}}
\vspace*{0.5cm}
\caption{ }
\end{figure}

\begin{figure}
\vspace*{-0.5cm}
\epsfxsize=3.8in
\centerline{\epsffile{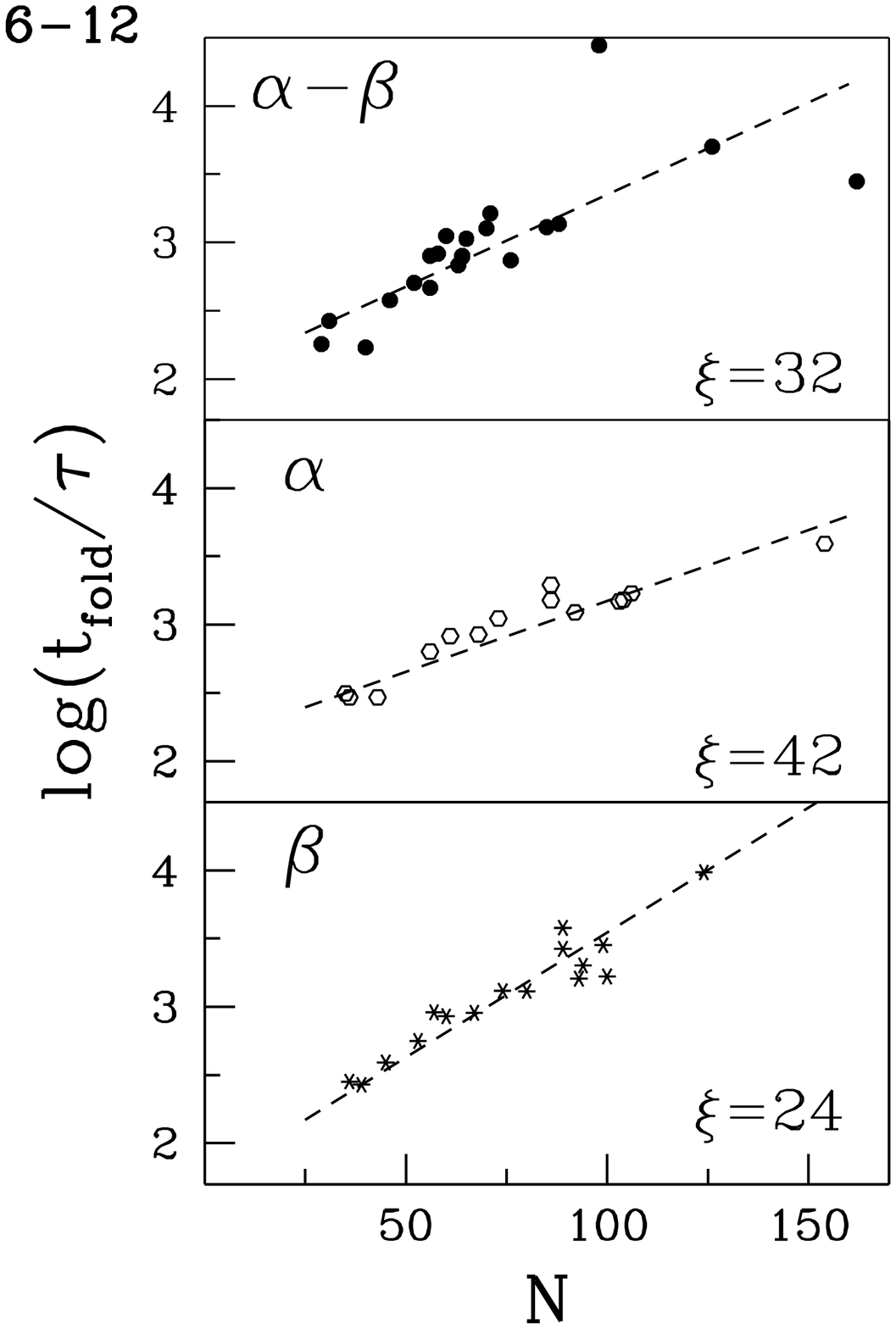}}
\vspace*{0.5cm}
\caption{ }
\end{figure}

\end{document}